\documentclass[graybox, envcountchap]{svmult}

\usepackage[]{natbib} 

\usepackage{mathptmx}        
\usepackage{amsmath}
\usepackage{amssymb}
\usepackage{color}
\usepackage{helvet}          
\usepackage{courier}         
\usepackage{dirtree}

\usepackage{makeidx}        
\usepackage{graphicx}        
\usepackage{subfig}

\usepackage{multicol}        
\usepackage[bottom]{footmisc}

\usepackage{hyperref}        
\hypersetup{colorlinks=true,urlcolor=blue}

\newcommand{\ml}{$ML$}
\newcommand{\infer}{$\mathcal{I}$}
\newcommand{\proppredict}{$\mathcal{P}_{t \to t+\Delta t}$}
\newcommand{\propcouple}{$\mathcal{P}_{1\to 2}$}
\newcommand{\emulate}{$\mathcal{E}$}

\providecommand{\Letter}{\textsuperscript{*}}

\makeindex             

\newcommand{\distributionstatement}[1]{%
  \begingroup
  \renewcommand{\thefootnote}{}%
  \footnotetext[0]{#1}%
  \endgroup
}

\begin{document}

\title{Machine learning applications in the heliosphere}
\author{Vishal Upendran and Banafsheh Ferdousi}
\institute{Vishal Upendran (\Letter) \at SETI Institute, Mountain View, CA, USA 94043; Lockheed Martin Solar and Astrophysics Laboratory, Palo Alto, CA, USA. \email{vupendran@seti.org}
\and Banafsheh Ferdousi\at Air Force Research Lab, Albuquerque, New Mexico, USA. \email{banafsheh.ferdousi@spaceforce.mil}}
%
\maketitle

\distributionstatement{DISTRIBUTION STATEMENT A. Approved for public release; distribution is unlimited. Public Affairs release approval AFRL20263729.}


\abstract{Space weather is the study of the near-Earth impact of the processes on the Sun. In a this era of technology, space weather emerges as a crucial area of study wherein the state of the near-Earth conditions need to be modeled and forecasted. Crucially, the advent of machine learning provides us with opportunities to study this subject in a rather comprehensive manner. In this chapter, we treat the the processes arising from Sun to the impact on the ground as a set of coupled systems, which can be studied in isolation or in a coupled manner. We define these individual systems in terms of their state vector $\vec{S}$, which provides us with an opportunity to define the physical variables determining the system. This also allows us to consider the systems in terms of their observables, obtained through the action of operators as $\vec{O}[\vec{S}]$. In such a formalism, the machine learning algorithms serve as {\it learned operators}, defined generally as inference ($\mathcal{I}$), propagation ($\mathcal{P}$), and emulation ($\mathcal{E}$). Through such a formalism, we present how machine learning helps advance our understanding of the physical systems under study, while also accelerating the development of operational models for space weather forecasting.}

\section{The Sun and heliosphere: A coupled, multiscale forecasting system}
\label{sec:spwml_intro}

The Sun profoundly impacts and modulates conditions in the interplanetary medium, including conditions on and near the Earth. The study of these variable plasma, magnetic, radiative, and energetic-particle conditions, and of their effects on technological and biological systems, is broadly called {\it space weather}. Space weather is extremely important from the perspective of technological infrastructure. One of the most beautiful visible manifestations of space weather is the aurora, the glow produced when charged particles precipitate into the upper atmosphere and excite its neutral constituents. However, space weather can also disturb radio communication, alter satellite drag, damage spacecraft electronics, degrade navigation systems, induce currents in long conductors on the ground, and disrupt electric power infrastructure, resulting in strong impact on humanity. In this chapter we view space weather not only as a collection of individual phenomena, but as a coupled heliophysical system whose state must be estimated, propagated, and forecast under severe observational and operational constraints.

The solar driver of space weather is intrinsically multiscale. On the longest relevant time scale, solar activity is organized by the approximately 11-year sunspot cycle and the associated 22-year magnetic polarity cycle. During solar minimum the large-scale magnetic field is closer to a dipole, coronal holes can produce recurrent high-speed solar-wind streams, and the heliosphere is often comparatively structured but persistent. During solar maximum the magnetic field becomes more complex, active regions emerge more frequently, and eruptive events become more common. On shorter time scales, active-region evolution, flares, coronal mass ejections (CMEs), stream interaction regions, shocks, and solar energetic particle (SEP) events inject energy and momentum into the corona and heliosphere over minutes to days. These drivers then propagate through the solar wind, interact with the terrestrial magnetosphere and ionosphere, and produce geoeffective responses on time scales ranging from seconds to several days.

This Sun-to-Earth chain may be described schematically as a sequence of coupled state vectors. The solar atmosphere supplies a source state, including magnetic field, plasma density, velocity, thermodynamic variables, radiation, and energetic-particle populations. The heliosphere evolves this state through magnetohydrodynamic transport, shocks, turbulence, magnetic connectivity, and particle acceleration. Near Earth, the upstream solar-wind and interplanetary magnetic-field state couples to the magnetosphere, ionosphere, thermosphere, and ground magnetic field. In this sense, space weather is a {\it system of systems}: the state vector of one subsystem becomes the boundary condition, forcing term, or latent driver of the next. Forecasting therefore requires not only recognizing signatures in individual data streams, but also learning how information is transformed and delayed as it moves across the heliosphere.

The observational data for this problem is very heterogeneous and multi-messenger. Remote-sensing instruments observe photospheric magnetic fields, chromospheric and coronal emission, coronagraphic CME structure, heliospheric-imager tracks, and radiative output across the electromagnetic spectrum. In-situ spacecraft measure the solar wind, magnetic field, plasma composition, energetic particles, and suprathermal populations at locations such as L1 and throughout the inner heliosphere. Ground networks measure magnetic perturbations, ionospheric total electron content, radio disturbances, and other impacts. These data products differ in cadence, latency, spatial coverage, calibration history, and physical proximity to the process being forecast. Some provide early but indirect information, such as photospheric magnetic predictors of flares or CMEs; others provide late but direct information, such as L1 solar-wind measurements that give tens of minutes of warning for geomagnetic response. A central challenge for machine learning is to combine these complementary information sources without confusing observational availability with physical causality.

Machine learning enters this setting as a family of inference and forecasting operators. Feature extraction, object detection, segmentation, representation learning, sequence modeling, and probabilistic classification can be viewed as operators that map observations into estimates of the hidden heliophysical state. Temporal models, surrogate models, and data-assimilation-inspired architectures then act as propagators that advance this state toward a forecast target. For example, a model may map magnetogram and extreme-ultraviolet observations to the probability of an M- or X-class flare in the next 24 hours or propagate upstream solar-wind measurements into global geomagnetic perturbation forecasts. These applications span classical machine-learning algorithms, deep convolutional and recurrent networks, transformers, graph representations, physics-informed surrogates, and hybrid empirical-physics systems~\cite{camporeale2018machine,2015ApJ...798..135B,2023NatSR..1313665A_operationalflare,2022FrASS...9.7301P,li_2025_wind_surrogate_coconut,2025SpWea..2304561B_cme_detect,2025arXiv251212786Y_sepforecast,2022SpWea..2003045U_geomag_fdl}.

Thus, we come to the domain of {\it operational forecasting}. The success of our defined operator is tied to action under uncertainty. A forecast has a target variable, an issue time, a valid time, a lead time, an update cadence, and a user-dependent cost function. The same physical event can require different products: a satellite operator may need SEP onset probability and radiation dose, a power-grid operator may need ground electric-field risk, an aviation user may need polar-cap absorption or radiation alerts, and a scientific modeler may need a calibrated distribution of solar-wind parameters. Thus the forecast object is not simply ``will an event occur?'', but a probabilistic statement about timing, magnitude, duration, location, and expected impact. This operational setting imposes constraints that are central to the design and evaluation of machine-learning models. Extreme events are rare, and the largest storms, fastest CMEs, and most intense SEP events are precisely the cases for which training examples are sparse. The data are nonstationary across the solar cycle, across instrument changes, and across evolving operational pipelines, including the impact of false alarms v/s missed events. Hence, the focus on a significant portion of this research rests on validation of models across time and activity. 

In this chapter we follow the Sun-to-Earth structure of the forecasting problem. We first discuss machine-learning methods for solar-wind and heliospheric-state estimation, where the objective is to infer and propagate the background medium through which transients travel (\S.~\ref{sec:solarwind}). We then turn to flare and eruptive-event forecasting, where the central difficulty is the connection between observable source-region evolution and rare energetic release (\S.~\ref{sec:flares}). We next discuss geomagnetic-storm prediction, where upstream solar-wind measurements, heliospheric context, and magnetospheric response must be converted into actionable forecasts (\S.~\ref{sec:magnetosphere}). Next, we consider near-Earth impacts in the ionosphere, thermosphere, and ground magnetic environment (\S.~\ref{sec:nearearth}). Finally, we shall conclude our review and speculate on future directions in \S.~\ref{sec:conclusion}.

\section{State vector formalism}
\label{sec:statevec}
We emphasize that the central object of interest, which sets the perspective on this chapter, is the state vector. The state vector can the set of density, velocity, magnetic field, ion populations and so on. However, for instance, in the in-situ case, the state vector may be the population distribution of different species, from which one can derive densities, momenta and so on. The state vector may be either almost directly be measured, or estimated through an observable. Thus, a state vector may look like $\vec{S} = \{\rho, T, B, ...\}$, where $\rho$ is the density, $T$ the temperature, $B$ the magnetic field, and so on.\\
From the previous chapter, we re-use the definition of some of the operators that act on the state vector. Specifically, these are:
\begin{svgraybox}
In this language, we define three major operator classes relating state vectors and observations across different systems:
\begin{eqnarray}
\mathcal{I}: \vec{O} = \mathcal{O}[\vec{S}] \rightarrow \hat{\vec{S}}, \quad
\mathcal{P}_{a\to b}: \vec{S}_a \rightarrow \vec{S}_b, \quad
\mathcal{E}: (\theta, \vec{S}_0) \rightarrow \vec{S}_{\rm sim}.
\end{eqnarray}
The first defined operator is inference {\infer}, which is the inverse transformation of observable synthesis from state vector. The second operator $\mathcal{P}_{a\to b}$, quantifies the coupling between two state vectors (\propcouple), or causal evolution of the individual state vector (\proppredict). The final operator is {\emulate}, and it essentially acts to serve as a surrogate methodology of a translating from a set of boundary / initial conditions to a full-volume representation.
\end{svgraybox}
We shall use this language to define the specific operations being performed on different state vectors. While these three operators are typically found in literature, we shall define new operators as and when applicable for different problems. This is a result of the complexity of the Sun-heliosphere-Earth system, which can be approached from several directions and in diverse manners. 
\section{Solar wind}
\label{sec:solarwind}

The solar wind is the persistent outflow of magnetized plasma from the solar corona into the heliosphere.It is a time-dependent plasma state either sampled by spacecraft or inferred through imaging observations, coronal and heliospheric models. The state vector of the solar wind may be written schematically as
\begin{equation}
    \vec{S}_{\rm sw}(\vec{x},t) =
    \{\rho, \vec{v}, \vec{B}, T, p, q_i, f_s, \ldots\},
    \label{eqn:sw_statevec}
\end{equation}
where $\rho$ is the mass density, $\vec{v}$ the bulk velocity, $\vec{B}$ the magnetic field, $T$ and $p$ the thermodynamic state, $q_i$ the composition and charge-state information, $f_s$ the particle distribution functions of different species, and these are a function of space ($\vec{x}$) and time ($t$). Routine measurements of the solar wind properties are generally performed in-situ, so these are single-point measurements in space which change in time and with the motion of the spacecraft. However, we also have approximate measurements of solar wind speed through Interplanetary Scintillation Measurements (IPS). These provide us with an approximate solar wind speed estimate across the heliosphere.

\subsection{Solar wind forecasting}
\label{subsec:sw_speed}
One of the most important aspects of modeling the heliosphere is to forecast the solar wind properties. Typically, this involves using the current state of the solar atmosphere, especially the corona ($\vec{S}_{\rm cor}(t_0)$) to predict the future state of the solar wind $ S_{\rm sw}(t_0+\Delta t)$.  The solar wind forecasting operator can hence be written down as:
\begin{equation}
    \mathcal{P}_{\rm cor \rightarrow sw} :
    \vec{S}_{\rm cor}(t_0) \rightarrow S_{\rm sw}(t_0+\Delta t).
    \label{eqn:cor_to_sw}
\end{equation}
From an operations perspective, routine solar wind measurements are available at L1, and the solar atmosphere is routinely observed only on the side facing the Earth. Typically, it is of interest to forecast the solar wind speed at L1 given changing conditions on the Sun. This target is physically motivated because speed controls stream interaction, solar-wind dynamic pressure, and the travel time of recurrent high-speed streams. However, we should note that the solar-wind speed is not a purely local quantity at Earth. It is shaped by the expansion of open magnetic flux tubes, coronal heating, stream-stream interactions, and the longitudinal mapping between the solar source region and the spacecraft.\\
Traditionally, the heliosphere is modeled as a {\it boundary value problem}, consisting of three steps: (i). Determining the state of the coronal magnetic field ($\vec{S}_{\rm cor}(t_0)$), (ii). Generating a boundary estimate of the solar wind ($ S_{\rm sw}(t_0)$), and (iii). Propagating the wind in the heliosphere ($ S_{\rm sw}(t_0+\Delta t)$). The global coronal magnetic field is estimated by assembling surface magnetic field maps across a full solar rotation, and extrapolating the field with the boundary. From this global field, empirical relations may provide velocity estimates by considering magnetic field topology properties. Finally, the propagation of solar wind may be accomplished by simplified 1D models, or through complex MHD simulations. The specific choice of free parameters for each component of this modeling scheme can be fine-tuned for specific periods of measurement, and for the measurement location in the interplanetary medium. Generally speaking, the kind of solar wind one measures depends on the source region for that particular solar wind, which is marked by wind speeds, plasma composition, magnetic field, etc. \\
{\ml} has been leveraged at almost every stage of this modeling scheme. For instance, the free parameters in the empirical relations in determining $S_{\rm sw}(t_0)$ have been optimized for each solar rotation using a neural optimizer, with transformer models generalizing the mapping across all rotations~\cite{sw_neural_wsa_2025}, eliminating the problem of unconstrained hyper-parameters. On the other hand, {\ml} methods have also been used to estimate $S_{\rm sw}(t_0+\Delta t)$ directly from $\vec{S}_{\rm cor}(t_0)$, allowing the network to  learn a nonlinear mapping from source-surface magnetic geometry to the later in-situ speed~\cite{sw_speed_coronal_field_2024,sw_pfss_cnn_2023}. \\
The step \#(ii) described above is an approximate empirical estimate of solar wind speed, calibrated through observations at L1. Instead of using this empirical relation, there has been effort in considering velocity estimates from IPS measurements to drive $S_{\rm sw}(t_0)\rightarrow S_{\rm sw}(t_0+\Delta t)$. However, IPS measurements are not available across all the times. To estimate the wind speed at all times, ANNs have been developed to map from $\vec{S}_{\rm cor}(t_0)\rightarrow S_{\rm sw}(t_0)$~\cite{2019SoPh..294..111Y,2021Univ....7..371Y_3dsolarwindreconstr}. 

The $\vec{S}_{\rm cor}(t_0)$ estimated through magnetogram modeling is not the most updated snapshot of the Sun, since it contains stale information assembled over several days. On the other hand, intensity images in EUV provide a high time cadence and nearly upto date estimate of $\vec{S}_{\rm cor}(t_0)$. This is possible since magnetic flux is mostly frozen in the corona, and the intensity approximately-traces the magnetic field topology. Hence, it is appealing to learn the specific mapping from intensity structures to solar wind measurements. {\ml} models spanning several kinds of architectures have been developed to summarize this time series of image data to forecast the wind speed several days ahead~\cite{2020_upendran,sw_ch_features_2023,sw_aia_3day_2023,sw_hss_images_2025,sw_hss_coronal_hole_2026}. The intention behind such modeling is for {\ml} to capture complex spatio-temporal associations between regions on the Sun and the measured wind type. However, typical deep learning models operate as black boxes, and the possible learning mechanism of the model remain hypothesis. To this end, the black box was opened up to quantify the solar wind sources by~\cite{2020_upendran}. This was one of the first works that demonstrated the association between coronal hole - fast solar wind and active region - slow solar wind is learned by a deep learning model. This was accomplished through the application of Grad-CAM on top of the model, and is quantified as a function of time and morphology through segmentation methods. Such work demonstrates that deep models inherently learn the known physics in an fully data-driven manner.

There is also a simpler time-series version of this task, where the input is a history of measured solar-wind quantities and the target is the future speed, effectively making this an auto-regressive task. These models are developed have been developed with the perspective of operational forecasting, since they are fast and rely on well-understood physics~\cite{sw_cnn_forecasting_2021,sw_gat_2022,sw_balnet_2025,sw_ensemble_lisa_2025,sw_multimodal_l1_2025}. In such cases, however, the model is effectively learning an approximation to $\mathcal{P}_{\Delta t}\vec{S}_{\rm sw}$, and depends on the smoothness of a one-dimensional time series.

\subsection{Surrogate models}
\label{subsec:sw_surrogates}
A second class of applications assumes that the empirical or global MHD models are correct, and attempt to replace the slow global model with fast {\ml} surrogate models. The learned operator can be written as
\begin{equation}
    \mathcal{E}_{\rm MHD} :
    \vec{S}_{\rm cor}(t) \rightarrow
    \hat{\vec{S}}_{\rm sw}(\vec{x},t+\Delta t),
    \label{eqn:sw_emulator}
\end{equation}
where $\mathcal{E}_{\rm MHD}$ is trained to emulate a numerical model, a reconstruction pipeline, or a data-assimilative heliospheric state estimate.

For instance, global MHD modeling can translate the photospheric magnetic field estimate $\vec{S}_{\rm phot}$ to coronal state estimation $\vec{S}_{\rm cor}$. The expensive MHD model can be replaced by an {\ml} surrogate made of U-nets, enabling the estimation to be sped up by a factor of $\sim1800$~\cite{li_2025_wind_surrogate_coconut}. One of the central innovations to learning this surrogate model is conceptualizing a rollout-loss, wherein $\vec{S}_{\rm sw}(r_i)$ feeds as an input to predict $\vec{S}_{\rm sw}(r_{i+1})$. Such a scheme allows loss estimation across several radii, and even enable a teacher-forcing methodology where the MHD target is fed as input during training time, allowing faster model convergence~\cite{li_2025_wind_surrogate_coconut,sw_spherical_fno_2025,sw_autoregressive_spherical_fno_2025}. These methods are attractive because they can provide global context at a cost closer to inference than simulation.

The advantage of a surrogate is speed, but the scientific risk is that the learned model may reproduce the training simulator rather than the heliosphere. This distinction is important. If the target is an MHD run, the neural network inherits the assumptions of that run: single-fluid closure, prescribed heating, simplified boundary conditions, limited turbulence, and numerical diffusion. A good surrogate should therefore be evaluated not only by pixel-wise or grid-wise error, but by whether it preserves physically important structures such as stream interfaces, rarefaction regions, latitudinal gradients, and the Parker-spiral magnetic geometry. In this sense, solar-wind surrogates are closer to reduced physical models than to generic regressors.

An adjacent direction is to drive heliospheric models directly with in-situ observations~\cite{sw_insitu_driven_2026}. Rather than relying entirely on a $\vec{S}_{\rm cor}$, the model incorporates spacecraft measurements as constraints on the evolving heliospheric state. In such methods, the in-situ time series $\vec{S}_{\rm sw}(t)$ updates the hidden state of the model $\vec{S}_{\rm cor}(t)$, while the model supplies spatial context that a single spacecraft cannot observe. This is especially important for the solar wind, where longitudinal structure and stream interaction make one-dimensional measurements ambiguous.

\subsection{Structure identification}
\label{subsec:sw_classification}
Solar-wind plasma is often divided into regimes such as coronal-hole wind, streamer-belt wind, sector-reversal plasma, slow Alfv\'enic wind, and compressed interaction regions. These labels are useful because they connect local measurements to global magnetic field topologies corresponding to source regions and transport histories. However, the boundaries between categories are not sharp surfaces in $\vec{S}_{\rm sw}$. They are overlapping regions in a multidimensional plasma-parameter space. Thus regime identification can be written as
\begin{equation}
    \mathcal{C}_{\rm sw} :
    \vec{O}_{\rm sc}(t-\tau:t) \rightarrow
    c_{\rm sw},
    \label{eqn:sw_classification}
\end{equation}
where $c_{\rm sw}$ is a discrete state label assigned to a time interval.

Gaussian-process methods have been used to classify the solar wind into multiple categories from in-situ data~\cite{sw_classification_gp_2017}. Similarly, {\ml} based categorization has been developed to separate solar-wind types and thereby support physical interpretation of wind evolution in interplanetary space~\cite{sw_categorization_2020}. A particularly important direction is objective determination of solar-wind states, where the goal is not to reproduce a human catalog exactly, but to discover reproducible plasma regimes that can be traced back to solar origins~\cite{sw_states_ml_2020}. More recent classification studies include uncertainty estimation, which is critical because ambiguous intervals should not be forced into a confident label~\cite{2024SoPh..299..131N_solar_wind_classify}. \\

Beyond broad regimes, the solar wind contains coherent structures across a wide range of scales. A structure in the solar wind may be defined as a subset of space-time for which the local plasma state satisfies a physical pattern,
\begin{equation}
    \mathcal{F}_{\rm sw} =
    \{(\vec{x},t) \mid \vec{S}_{\rm sw}(\vec{x},t) \in
    \Omega_{\rm struct}\},
    \label{eqn:sw_structure}
\end{equation}
where $\Omega_{\rm struct}$ is the region of state space corresponding to a structure class. Such a state space may correspond to, for instance, shock boundaries in the streaming plasma. Since a spacecraft samples only a trajectory through $\mathcal{F}_{\rm sw}$, the practical problem becomes detection from multivariate time series.

{\ml} has been successfully applied to the detection of magnetic flux ropes in this spirit. Using existing datasets, simple algorithms like Random forests are able to discern flux rope signatures in solar wind time series~\cite{sw_flux_rope_properties_2024}. A more pertinent issue arises due to the 3-D structure of the flux rope, and a lossy 1-D sampling by in-situ measurements. However, the time series of the sampled B field can be used to get a general estimate of the orientation of the 3D structure. This is seen through simulation-based inference, wherein a CNN is able to ingest the time series of B field measurements to get a general sense of the 3D orientation of the flux rope~\cite{sw_flux_rope_orientation_2022}. Interplanetary shocks and discontinuities provide another example of structure identification. Automatic shock identification can be posed as a classification problem on local windows of plasma and magnetic-field data, and simple MLPs are able to solve this problem well~\cite{sw_interplanetary_shocks_2026}. Apart from shocked structures, we also see periodic structures like waves. CNNs have proved effective in differentiating non-periodic structures from periodic structures, by leveraging convolution using kernels of multiple sizes in parallel, similar to the inception module~\cite{swstruct_2023ApJ...949...40F}. Such efforts enable us to trawl through massive datasets, and generate statistics on events of interest, which presents us with a better understanding of the solar wind.
\section{Flares, CMEs, and solar energetic particles}
\label{sec:flares}
\begin{figure}
    \centering
    \includegraphics[width=\linewidth]{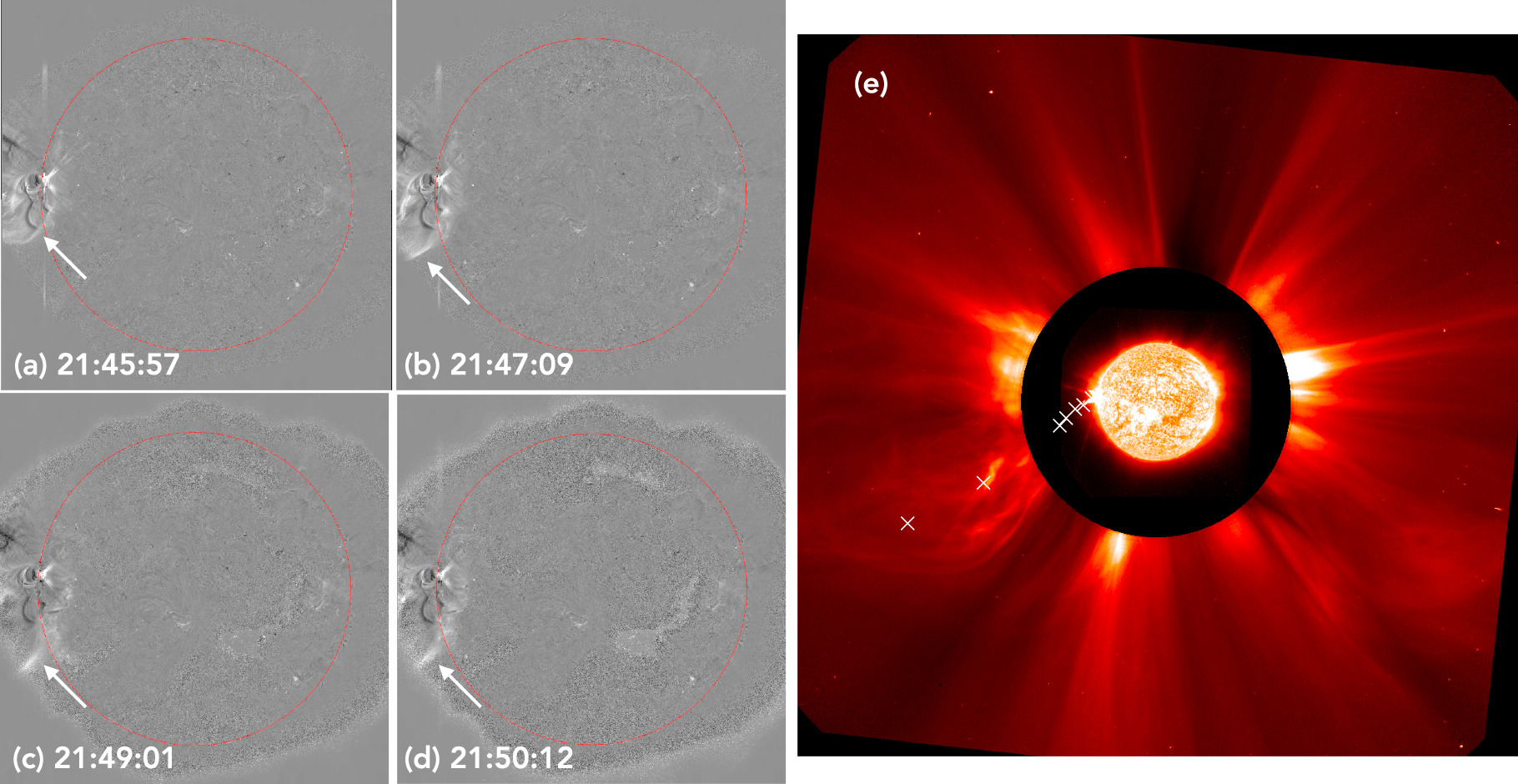}
    \caption{Panels (a)-–(d) display a sequence of base difference images with respect to 21:30:57 UT in AIA 211~{\AA} showing the propagation of the EUV wave that originated from a solar flare. White arrows mark the EUV wave. Panel (e) shows the SUVI 304~{\AA} and LASCO/C2 observation of the CME originating from the flare. Figure taken from~\cite{cme_figure}.}
    \label{fig:cmefig}
\end{figure}

Solar eruptive activity is one of the most impulsive and dynamic part of the Sun-to-Earth system. Magnetic free energy stored in active regions is suddenly released, and is used to heat up plasma locally, accelerate particles, and cause bulk plasma motion. The release of energy may also launch large scale magnetic field structures into the heliosphere. Such events are manifested as solar flares on the Sun, and as ccoronal mass ejections (CMEs) \& solar energetic particles (SEPs) in the heliosphere. Generally, these phenomena are all related, but may occur without any visible signature of the other. In Fig.~\ref{fig:cmefig}, we display the result of a flare on the whole of the observed Sun. Panels (a) -- (d) show the propagation of a global perturbation arising from the flare. This flare emits a flux rope that is tracked later in coronograph data in panel (e), as depicted by the white white crosses. This image is to scale -- the reader may note the spatial scale varies over multiple orders of magnitude on traveling through space, thus illuminating the complexity and coupling of the different systems. \\
The state vector of the atmosphere responsible for the eruptive event can be defined as:
\begin{equation}
    \vec{S}_{\rm \odot}(t) =
    \{\vec{B}_{\rm phot}, \vec{B}_{\rm cor}, \rho, T,
    \vec{v}, E_{\rm free}, H, \ldots\},
    \label{eqn:eruptive_statevec}
\end{equation}
where $\vec{B}_{\rm phot}$ and $\vec{B}_{\rm cor}$ describe the photospheric and coronal magnetic field, $\rho$ represents density,  $T$ would be temperature, $E_{\rm free}$ and $H$ denote magnetic free energy and helicity-like quantities. In general, this state vector may not be very well determined in the first place. For instance, the specific physical variables that trigger solar flares are still very elusive. Starting out from this state of the atmosphere $\vec{S}_{\rm atmos}$, one can define the forecasting problem as:
\begin{equation}
    \mathcal{P}_{\rm erup} :
    \vec{S}_{\rm \odot}(t_0-\tau:t_0)
    \rightarrow
    p(y_{\rm flare}, y_{\rm CME}, y_{\rm SEP}, t_{\rm onset} \mid t_0),
    \label{eqn:eruptive_operator}
\end{equation}
where  $t_0$ is a reference time, $\tau$ corresponds to the time history of the Sun taken into consideration, $y_{\rm flare}$, $y_{\rm CME}$, and $y_{\rm SEP}$ are event labels, and $t_{\rm onset}$ is the onset time. In practice, however, one does not observe the state vector of physical variables, resulting in the presence of only the observable $\vec{O}[\vec{S}_\odot]$. Hence, more generally, the forecasting problem can be modified to be written as 
\begin{equation}
    \mathcal{P}_{\rm erup} :
    \vec{O}[\vec{S_\odot}](t_0-\tau:t_0)
    \rightarrow
    p(y_{\rm flare}, y_{\rm CME}, y_{\rm SEP}, t_{\rm onset} \mid t_0).
    \label{eqn:eruptive_operator_obs}
\end{equation}
The central difficulty of these class of problems is that the most important cases are rare and often lie near the edge of the observed distribution. Hence, it becomes extremely important to have the right metrics of success, calibration across data series and across the solar cycle, and the right definition of events.
\subsection{Solar flare}
\label{subsec:flare_forecasting}
Solar flare forecasting is one of the most extensive machine-learning applications in heliophysics~\cite{2015ApJ...798..135B}. A common task is to predict whether an active region will produce a flare above a given GOES class threshold within a fixed window, for example an M- or X-class flare within the next 24~hr. This can be written as
\begin{equation}
    \mathcal{C}_{\rm flare} :
    \vec{O} [\vec{S}_{\rm AR}](t_0-\tau:t_0)
    \rightarrow p(F_{\geq M}(t_0,t_0+\Delta t)),
    \label{eqn:flare_classifier}
\end{equation}
where $\vec{O} [\vec{S}_{\rm AR}]$ is the active-region observation history and $F_{\geq M}$ is the event that the region produces at least one flare above the chosen threshold ($M$).

Early and still influential approaches use physically motivated active-region features. These include calculating quantities like total unsigned flux, magnetic gradients, shear, measures of non-potentiality, polarity inversion line properties, and temporal changes in magnetic complexity. Sequential supervised learning, magnetic non-potentiality, and magnetogram-based predictors all represent this feature-space view of the problem~\cite{2009SoPh..255...91Y,2013ApJ...774L..27Y,2015ApJ...798..135B,2018SoPh..293...28F_flarepredict,2017ApJ...849..148S}. The scientific value of these models is that the input variables are very interpretable. However, such approaches are limited by the intuition of the `important variable' as defined by the forecaster, and may not be able to capture the inherent complexity of the AR.\\
Deep-learning models relax this compression by operating directly on magnetograms or image-derived representations. Convolutional neural networks have been used for line-of-sight magnetograms and for flaring versus non-flaring active-region classification~\cite{2018ApJ...856....7H,2020ApJ...898...98B,2021AdSpR..67.2544A}. Transformer-based operational frameworks and comparative deep-learning studies extend this idea to sequence modeling and larger-scale training sets~\cite{2023NatSR..1313665A_operationalflare,2025ApJS..279...46L}. In this setting the model attempts to summarize morphology, gradients, polarity-inversion-line structure, and their temporal evolution into a predictive representation.\\
Since solar flares are inherently outside of the main distribution of $\vec{S}_{\odot}$, the extreme events are the most devastating and can have the largest impact. This outlier-effect has been attempted to be mitigated through several ways. One way is to have a class-dependent reward function that penalizes the developed model according to the class imbalance~\cite{2025ApJS..279...46L,2026MNRAS.547ag349W}. Another way is to consider more features in  $\vec{S}_{\odot}$ that enable maximal separation across different classes~\citep{hu_2025_Bfieldheight}. An altogether different trajectory is to move on from deterministic labels to calibrated probabilities and event times. Reliable probability forecasting explicitly asks whether a predicted probability can be interpreted as an event frequency~\cite{2020ApJ...899..150N}. Deep survival analysis recasts flare forecasting as a time-to-event problem rather than only a fixed-window classification problem~\cite{meyer_2025_flareforecast_survivalanalysis}. This is operationally meaningful because users care about both probability and timing. Such deep survival analysis provides a time-stratified forecasting probability, which is not the case for a fixed-time window forecast, where the flare can be immediate or concentrated near the end of the window.\\
The choice of input representation determines which part of $\vec{S}_{\rm AR}$ the model can access. Magnetograms emphasize the photospheric magnetic boundary. UV/EUV data provides information about the solar atmospheric thermodynamics and emission morphology.  Recent work therefore moves beyond a single static magnetogram toward multimodal and temporally structured inputs~\cite{Rosales_2026_flarepredict,vural_2025_multivarflare,2023ApJS..266....8S_euv_features}. Incorporating multimodal data enhances the predictability of flare, as there is a more comprehensive access to $\vec{S}_{\rm AR}$ when compared to single-modality cases. Furthermore, using a time series of data enables the model to learn the general temporal dynamics of ARs, and this again informs the onset of instability for the flare.\\
There is also a growing effort to make flare models diagnostically useful rather than only predictive. Visual explanations have been used to connect deep flare forecasts to physical active-region parameters~\cite{2021ApJ...910....8Y_interpretflare}. Full-disk context is another unresolved issue. Active-region models are natural when the source region is already identified, but operational systems must also recognize new emergence, multiple active regions, far-side evolution projected onto the visible disk, and events outside preselected cutouts. Coupling full-disk and active-region-based forecasting is one route toward this broader operational setting~\cite{2022FrASS...9.7301P}. This also highlights a subtle validation problem: a model may perform well when evaluated on curated active-region samples, but degrade when the event-detection, region-association, and forecasting problems are coupled together.

\subsection{Coronal Mass Ejections}
\label{subsec:cme_forecasting}
CMEs represent a level of complication above typical solar flares. The relevant operator is not only whether an eruption occurs, but also the probability of a launch of a coherent magnetized plasma structure, its trajectory, orientation, speed, angular width, and magnetic configuration. The source-region operator may be written as
\begin{equation}
    \mathcal{P}_{\rm CME,src} :
    \vec{O}[\vec{S}_{\rm AR}](t_0-\tau:t_0)
    \rightarrow p(y_{\rm CME}, \vec{S}_{\rm CME}(t_{\rm launch}) \mid t_0),
    \label{eqn:cme_source_operator}
\end{equation}
where $y_{\rm CME}$ is whether the CME is launched towards the Earth or not, and $\vec{S}_{\rm CME}$ contains properties such as launch time, speed, direction, width, and magnetic structure. 
Sometimes, the source regions of CMEs may not be well-seen in coronograph images, resulting in a halo-CME. Such CMEs may identified only near Earth, and not at the Sun, and this might be too late from an operational perspective. However, machines may pick up signatures that human eyes may not be able to perceive. In such a case, {\ml} has been used to predict halo CME occurrence from SDO/HMI vector magnetic data with transformer models~\cite{zhang_2025_cmeforecasting}. Such transformer models are able to capture temporal association between vector magnetic field data and whether a halo CME is produced or not. 

Not all flares result in CMEs, and a distinction needs to be made on which flares can actually give rise to CMEs. Standard active region summary properties have been used to train shallow {\ml} models to classify them as having only a flare or a flare + CME~\cite{cmeforecast_2008SoPh..248..471Q, cmeforecast_2016ApJ...821..127B}. These results indicate that the amount of horizontal shear and the amount of free energy in the system are strong indicators of whether the flare is accompanied by a CME or not. However, these methods still rely on the intuition of the observer to define the physical variables quantifying $\vec{O}[\vec{S}_{\rm AR}]$. This is mitigated at some level by multimodal deep models~\cite{cme_soho_2024SoPh..299..159A, cmesoho_2026SoPh..301...73Y}. Apart from showing good performance, these models are also opened up using visualization techniques like Integrated Gradients. Such analysis demonstrates that the model considers loops structures in the active region of interest for predicting a correct classification of the active region as giving rise to a CME.

Once a CME is launched, it travels outwards into the interplanetary medium. This, one is concerned with the propagation of the CME defined as:
\begin{equation}
    \mathcal{P}_{t\rightarrow t+\delta t} :
    \vec{S}_{\rm CME}(t)\rightarrow
    \vec{S}_{\rm CME}(t+\delta t),
    \label{eqn:cme_prop_operator}
\end{equation}
where we are propagating the state vector of the CME ($\vec{S}_{\rm CME}(t)$) in time. There is some amount of modeling in this domain, called drag-based models. These models assume a physical structure of the CME and propagate them through the solar wind using a simplified momentum equation. {\ml} has been incredible in such physics based modeling by learning the optimal coefficients of the equations from the data~\cite{cme_drag_2023ApJ...954..151G}. By training on historical data and using simplified physics as a constraint for the CME travel, such methods are able to provide travel times for even very complex scenario where two CMEs interact with each other on the way to the Earth~\cite{cme_drag_2026SoPh..301...53G}. Another way of estimating the arrival time is to learn the mapping from CME and solar wind parameters to previously identified travel times, using simpler {\ml} models~\cite{catpuma_2018ApJ...855..109L,catpuma_cmearrival_2024ApJ...963..121C}. Such methods are able to present a ranking of the most important parameters for arrival time prediction. The CME arrival time is found to be most influenced by its initial launch velocity, and not quite dependent on the background solar wind speed or density. While such a result may hold statistically since a CME is essentially a huge shock in the system, the scenario is very different with multiple interacting CMEs. 

The subject of CME prediction rests on very good datasets where the reported parameters are well representative of all situations possible. Hence, while {\ml} has been starting to help in this complex topic, care is needed to ensure the statistical results respond well to apparent pathological cases that are missed simply because they do not interact with the Earth.

\subsection{SEP forecasting}
\label{subsec:sep_forecasting}
SEP forecasting is the most explicitly coupled of the eruptive-event problems. The target is an energetic-particle flux or fluence at a spacecraft or planetary environment, but the source information comes from flares, CMEs, shocks, radio bursts, the ambient solar wind, and magnetic connectivity. A schematic SEP operator is
\begin{equation}
    \mathcal{P}_{\rm SEP} :
    \{\vec{O}[\vec{S}_{\rm flare}], \vec{O}[\vec{S}_{\rm CME}], \vec{S}_{\rm sw}, \vec{B}_{\rm IMF}\}
    \rightarrow p(J(E,t) > J_{\rm th}),
    \label{eqn:sep_operator}
\end{equation}
where $J(E,t)$ is the particle intensity at energy $E$ and $J_{\rm th}$ is an operational threshold. This target differs from flare and CME prediction because the event observed near Earth is shaped by both acceleration and transport.

SEP forecasting is a very mature domain of heliophysics, resulting in an extensive development of tools, data and metrics of evaluation. Typically, a hand crafted set of features comprising of both in-situ and remote sensing variables is considered. These variables are similar to the ones considered for the flares (\S.~\ref{subsec:flare_forecasting}) and CMEs (\S.~\ref{subsec:cme_forecasting}). The community has typically used shallow {\ml} models for SEP forecasting~\cite[see, for instance][]{sep_forecast_2021SoPh..296..107L,sep_forecast_2024ApJ...974..131K, 2025arXiv251212786Y_sepforecast,sep_forecast_2025ApJS..277...34H}, with limited deep learning applications. We refer to an extensive review of the subject, comparing datasets, data definitions, models and performance metrics may be found in ~\cite{sep_report_whitman2026validation}. \\
The predominant direction of the community has been not just performing model development, but developing evaluation strategies. The community has developed extensive evaluation datasets through the Community Coordinated Modeling Center~\footnote{\url{https://ccmc.gsfc.nasa.gov/scoreboards/sep/}}. The community has also incorporated as many modeling efforts as possible to present an approximately uniform evaluation strategy, which has great value for operators and stakeholders~\cite{sep_report_whitman2026validation}.


\section{The magnetosphere: Coupling, dynamics, and forecasting under solar wind forcing}
\label{sec:magnetosphere}
\begin{svgraybox}
The magnetosphere is the cavity formed where the geomagnetic field stands as an obstacle to the solar wind: the incoming flow is decelerated at the bow shock, deflected around the magnetopause on the dayside, and drawn out into an extended magnetotail on the nightside. Within this cavity, the geomagnetic field confines plasma populations of both solar wind and ionospheric origin. The magnetosphere is not a static structure but a dynamically coupled system whose configuration, particle populations, and electrodynamic state are continuously driven by the upstream solar wind and mutually coupled, through field-aligned currents, to the ionosphere and thermosphere. Its state vector may be written schematically as
\begin{equation}
    \vec{S}_{\rm mag}(\vec{x},t) =
    \{\vec{B},\, n_e,\, n_i,\, T_e,\, T_i,\,
      f_s(E,\alpha,L,\text{MLT}),\,
      P_{\rm rc},\, P_{\rm wave},\, \Phi_{\rm pc},\, \ldots\},
    \label{eqn:mag_statevec}
\end{equation}
where $\vec{B}$ is the global magnetic field configuration of Earth, $n_{e,i}$ and $T_{e,i}$ are the electron and ion densities and temperatures across the plasma regions (magnetosheath, plasma sheet, ring current, plasmasphere), $f_s(E,\alpha,L,\text{MLT})$ is the phase space distribution function of species $s$ as a function of energy $E$, pitch angle $\alpha$, L-shell, and magnetic local time, $P_{\rm rc}$ is the ring current pressure, $P_{\rm wave}$ is the plasma wave energy density (chorus, EMIC, hiss, magnetosonic, ULF), and $\Phi_{\rm pc}$ is the cross-polar-cap potential. This state spans spatial scales from the electron gyroradius to the $>$100~Earth radius ($R_E$) magnetotail, and time scales from wave-particle interaction timescales of seconds to storm recovery timescales of days.
\end{svgraybox}

A central observational difficulty is that $\vec{S}_{\rm mag}$ is never directly observed in full. Spacecraft sample only a trajectory through the system, providing in-situ measurements at a limited set of locations. Ground-based magnetometer networks sample the magnetic imprint of large-scale current systems at the Earth's surface. Remote sensing provides auroral images that encode the energy deposition pattern of precipitating particles. The observed quantities are therefore always a compressed, partial projection of the full state:
\begin{equation}
    \vec{O}[\vec{S}_{\rm mag}] =
    \{\delta\vec{B}_{\rm gnd},\,
      J_{\rm FAC},\,
      \mathcal{I}_{\rm aurora},\,
      J_{\rm sc}(E,t)\,
      \ldots\},
    \label{eqn:mag_obs}
\end{equation}
where $\delta\vec{B}_{\rm gnd}$ is the ground magnetic perturbation, $J_{\rm FAC}$ the field-aligned current density inferred from low-Earth-orbit satellite constellations, $\mathcal{I}_{\rm aurora}$ the auroral intensity distribution, and $J_{\rm sc}(E,t)$ the in-situ particle flux at spacecraft locations. The upstream solar wind state $\vec{S}_{\rm sw}^{\rm L1}$ measured at L1 serves as its primary external boundary condition. Observers and systems operators monitor geomagnetic activity through indices like Kp, Dst, SYM-H, and AE, which are quantities derived from summary of $\vec{O}[\vec{S}_{\rm mag}]$, and are designed to compress the magnetospheric state into operationally useful scalars. {\ml} therefore operates on these observation rather than on $\vec{S}_{\rm mag}$ itself -- and its task is to learn the operators that connect the solar wind forcing at the boundary to the particle populations, current systems, and wave environments deep inside the magnetosphere.

\subsection{Solar wind--magnetosphere coupling}
\label{subsec:sw_mag_coupling}

The fundamental driver of the magnetospheric state is the solar wind. The coupling operator
\begin{equation}
    \mathcal{P}_{\rm sw \to mag}:
    \vec{S}_{\rm sw}(t_0-\tau:t_0)
    \rightarrow
    \vec{S}_{\rm mag}(\vec{x},\, t_0+\Delta t),
    \label{eqn:sw_mag_propagator}
\end{equation}
maps a solar wind time history over window $\tau$ to the magnetospheric state at lead time $\Delta t$. Energy and mass cross the magnetopause through several distinct mechanisms. The dominant channel during southward IMF is dayside reconnection, which opens field lines to the solar wind and drives the Dungey convection cycle~\cite{1961PhRvL...6...47D}. During northward or variable IMF the Kelvin-Helmholtz instability (KHI) along the magnetopause flanks drives vortex-mediated entrainment of solar wind plasma~\cite{2015NatCo...6.7019K, 2023NatCo..14.2513K}, and flux transfer events (FTEs) provide a third pathway through transient reconnection bursts that create short-lived flux tubes threading the boundary~\cite{1979GeoRL...6...33R}. Additional mechanisms -- viscous diffusion, impulsive penetration, and high-latitude lobe reconnection -- also contribute, with their relative importance depending on IMF orientation, dynamic pressure, and Mach number. Empirical scalar functions such as the Newell coupling parameter~\cite{2007JGRA..112.1206N} collapse this multi-mechanism interaction into a single measure, inevitably omitting information in the full high-dimensional input space.

The most directly observable consequence of Eq.~\ref{eqn:sw_mag_propagator} is the configuration of the magnetospheric boundary, and ML approaches to it address two complementary tasks: characterizing the physical conditions at and upstream of the magnetopause, and identifying which plasma region a spacecraft occupies. For boundary characterization, ML models trained on multi-spacecraft data recover magnetopause standoff distance and shape from solar wind dynamic pressure and IMF orientation~\cite{2025AnGeo..43..835K}; a probabilistic neural network replaces the conventional planar propagation method for the L1-to-magnetopause solar wind transit~\cite{2023FrASS..1050779O}, and a companion probabilistic RNN characterizes the resulting magnetosheath conditions from upstream L1 inputs~\cite{2024JGRH....1..235O}. For region identification, unsupervised clustering of eight years of MMS dayside crossings maps the solar wind, magnetosheath, and magnetosphere boundaries without hand-labeled catalogs~\cite{2024JGRA..12932431T}; supervised CNN plus Random Forest classification extends this to five plasma environments at finer regional resolution~\cite{2025arXiv251000190A}. Transfer learning of these region-identification operators generalizes across planetary environments~\cite{2025FrASS..1208091T}.

A complementary view of $\mathcal{P}_{\rm sw \to mag}$ comes from data-driven modeling and decomposition. ML decomposition of simultaneous solar wind and magnetotail measurements reveals the physical structure within this operator, identifying dual energy and mass transfer channels -- one driven by the solar wind electric field and one by ram pressure~\cite{2026JGRA..13134883Z}. The resulting output state space is mapped by clustering multi-spacecraft THEMIS observations: the magnetosphere organizes into reproducible plasma regimes with systematic solar wind dependence, providing both an empirical atlas and a benchmark for surrogate evaluation~\cite{2024JGRH....1..221E}. At the global simulation level, graph-based neural network surrogates of hybrid-Vlasov codes have demonstrated that the spatiotemporal evolution of electromagnetic fields can be reproduced at a fraction of the simulation cost, extending fast emulation to kinetic regimes inaccessible to standard MHD~\cite{2026arXiv260112614H}.

A key scalar target for ML studies of $\mathcal{P}_{\rm sw \to mag}$ is the cross-polar-cap potential ($\Phi_{\rm pc}$, CPCP) -- the total electric potential drop across the polar cap driven by the solar wind electric field, and the primary scalar linking SW-M coupling to ionospheric electrodynamics. The mapping
\begin{equation}
    \mathcal{P}_{\rm sw \to \Phi}:
    \{v_{\rm sw},\, B_{\rm IMF},\, \theta_{\rm clock}\}(t-\tau:t)
    \rightarrow
    \Phi_{\rm pc}(t+\Delta t)
    \label{eqn:cpcp}
\end{equation}
has long been thought to be nonlinear, with $\Phi_{\rm pc}$ appearing to saturate above $E_m^* \approx 10$--$12$~mV~m$^{-1}$ of the solar wind merging electric field $E_m^* = v_{\rm sw} B_T \sin^2(\theta_{\rm clock}/2)$, spawning more than ten competing physical theories. A new study~\cite{2026Natur.Sivadas} has demonstrated that this saturation is a statistical artifact: $E_m^*$ measured at L1 is a heteroskedastic proxy for the true near-reconnection-site field $E_m^{\rm sh}$, with relative uncertainty of at least 30\% that grows with solar wind strength; regression calibration correcting this bias reveals a linear geomagnetic response throughout and suggests extreme storm impacts may be up to twice as large as previously inferred. This poses a direct and unresolved challenge for ML: models trained on L1 data risk learning the measurement bias as physical structure, with the artifact likely present in widely used ML outputs across space physics. SuperDARN-based analyses that captured apparent saturation~\cite{2019JASTP.19305057L} should therefore be reinterpreted as learning regression-to-the-mean bias; neural network emulators of MHD simulations~\cite{2023EP&S...75..139K} and data-driven ANNs trained on satellite observations~\cite{2025E&SS...1204331L} inherit the same caveat: any nonlinearity under strong driving in the learned $\mathcal{P}_{\rm sw \to \Phi}$ operator may mirror the bias of erroneous L1 inputs.

\subsection{Geomagnetic indices as compressed observables}
\label{subsec:geoindices}

Because $\vec{S}_{\rm mag}$ is not directly observable, the community has historically summarized the magnetospheric state through scalar indices derived from ground magnetometer networks. These are inference operators applied to $\vec{O}[\vec{S}_{\rm mag}]$:
\begin{equation}
    \mathcal{I}_{\rm idx}:
    \vec{O}[\vec{S}_{\rm mag}](t)
    \rightarrow
    \{K_p,\, \text{Dst},\, \text{SYM-H},\, \text{AE},\, \text{AL}\}(t).
    \label{eqn:index_inference}
\end{equation}
The $K_p$ index is a semi-logarithmic measure of the maximum ground magnetic perturbation over three-hour intervals at mid-latitude stations, and it serves as the primary global activity indicator in space weather operations. The Dst index measures the symmetric horizontal component of the magnetic field at equatorial stations and is a proxy for the ring current pressure $P_{\rm rc}$: a negative Dst excursion reflects the diamagnetic field depression produced by the injection and energization of ions into the ring current during a geomagnetic storm. SYM-H is a one-minute-resolution analogue of Dst. The AE (auroral electrojet) and AL indices measure the strength of the high-latitude electrojet current system and are sensitive to substorm activity.

A critical point for machine learning is that these indices are compressed projections, not the state itself. A model that predicts $K_p$ or Dst is not modeling the magnetosphere -- it is modeling the end-to-end composition of $\mathcal{P}_{\rm sw \to mag}$ and $\mathcal{I}_{\rm idx}$, short-circuiting the physical intermediate states. This is operationally valuable but scientifically insufficient: the same Dst value can correspond to very different spatial distributions of ring current pressure, different particle compositions, and different upstream drivers. Interpreting ML model weights or attention maps in terms of physical mechanisms must account for this compositionality.

ML approaches for geomagnetic index prediction address the propagation operator
\begin{equation}
    \mathcal{P}_{\rm sw \to idx}:
    \{v_{\rm sw},\, n_{\rm sw},\, B_{\rm IMF},\, B_z^{\rm IMF}\}
    (t-\tau:t)
    \rightarrow
    \text{idx}(t+\Delta t),
    \label{eqn:idx_propagator}
\end{equation}
where the lead time $\Delta t$ ranges from minutes (nowcasting) to 72 hours (operational planning). Recurrent architectures such as LSTMs have become standard because the magnetospheric response has memory: ring current decay, plasmaspheric refilling, and substorm recurrence impose temporal structure on $\text{idx}(t)$ that depends on the solar wind history $\vec{S}_{\rm sw}(t-\tau:t)$. Feature selection studies have established which solar wind parameters carry independent predictive information for Dst beyond $B_z$ and velocity, identifying solar wind density and the dynamic pressure impulse as key contributors to main-phase injection~\cite{2022ApJS..260....6L}. LSTMs with attention mechanisms applied to $K_p$ forecasting have demonstrated that the attention weights preferentially focus on IMF $B_z$ reversals and solar wind speed enhancements, recovering the known physics of magnetic reconnection driving without being explicitly trained on it~\cite{2026JASTP.28406839K}. Multi-fidelity boosted neural networks provide a complementary path, combining models trained on different data fidelities to deliver calibrated multi-hour-ahead Dst forecasts~\cite{2023SpWea..2103286H}. Probabilistic and multi-index formulations extend this to joint forecasts of $K_p$, Dst, SYM-H, and AE simultaneously, capturing cross-index correlations that reflect coupled storm-substorm dynamics~\cite{2026arXiv260614417Y}. Uncertainty quantification has emerged as a critical addition: a single deterministic forecast of Dst cannot convey whether the model is confident or extrapolating, and ensemble or Bayesian approaches provide calibrated probability distributions over future storm severity~\cite{2024SpWea..2204002W}.

A broader challenge is that index prediction models trained on moderate activity inevitably underperform during the largest storms, which lie outside the training distribution; calibration must therefore be evaluated not only on aggregate skill scores but on conditional performance at high-$K_p$ thresholds~\cite{2025GeoRL..5214848K}. Addressing this imbalance at the training stage -- by pruning uninformative moderate-activity samples to improve performance on rare extreme events -- provides a principled complement to evaluation-focused approaches~\cite{2025arXiv251206950C}.

\subsection{Storm-time ring current and inner magnetosphere}
\label{subsec:ring_current}

The ring current is the population of energetic ions (predominantly O$^+$ and H$^+$ at 10--300 keV) trapped on closed drift paths encircling the Earth at geocentric distances of $2$--$7\,R_E$~\cite{1999RvGeo..37..407D}. It is the dominant source of the negative Dst perturbation during geomagnetic storms, and its state is governed by the Liouville equation for the ion phase space distribution:
\begin{equation}
    \frac{\partial f_i}{\partial t} +
    \{\mathcal{H}, f_i\} =
    Q(E,\alpha,L) - L(E,\alpha,L,f_i),
    \label{eqn:ring_current_liouville}
\end{equation}
where $\{\mathcal{H}, f_i\}$ is the Poisson bracket encoding drift motion under the guiding-center Hamiltonian, $Q$ is a source (injection from the plasma sheet driven by electric fields during the storm main phase), and $L$ is a loss term (charge exchange with neutrals, wave-particle scattering). The ring current state vector is therefore $f_i(E,\alpha,L,\text{MLT},t)$, and its projection onto Dst via the Dessler-Parker-Sckopke relation provides the scalar observable:
\begin{equation}
    \text{Dst}(t) \approx -\frac{\mu_0}{2\pi B_0 R_E^3}
    \int \!\! E_\perp \, f_i \, dE \, d\alpha \, dL \, d\text{MLT},
    \label{eqn:dps_relation}
\end{equation}
where $B_0$ is the equatorial surface field and $E_\perp$ is the perpendicular energy. Physics-based codes such as RAM-SCB propagate $f_i$ forward using electric and magnetic field models as input; they are accurate but computationally expensive, precluding ensemble or real-time applications.

ML emulation of ring current physics-based models provides a fast surrogate that preserves the spatial structure of the distribution. A reduced-order probabilistic emulator of RAM-SCB has been demonstrated to reproduce storm-time ring current pressure distributions and Dst traces at a small fraction of the computational cost, with calibrated uncertainty arising from uncertainty in the electric field input~\cite{2024SpWea..2203706C}. Similarly, neural network models for the spatial distribution of ring current proton and oxygen populations reproduce the dawn-dusk asymmetry and species-dependent injection patterns characteristic of different storm intensities~\cite{2025FrASS..1229056L, 2024SpWea..2203779W}. Beyond surrogate modeling, ML has also been used to reconstruct the ring current state during historical periods predating modern in-situ coverage: by applying the learned mapping to reconstructed solar wind inputs, the ring current response to pre-satellite extreme events can be estimated~\cite{2025SpWea..2304293S}. The equatorial electric field driving this injection has itself been modeled with neural networks, recovering the convection pattern from upstream conditions and providing an ML-based specification of the convection-driven transport channel~\cite{2026GeoRL..5320163H}.

\subsection{Radiation belts and wave-particle interactions}
\label{subsec:rad_belts}

The radiation belts consist of electrons and protons trapped on bounce and drift trajectories at L-shells of $1.2$ to $\sim7\,R_E$, with energies ranging from hundreds of keV to tens of MeV. Their dynamics are governed by the Fokker-Planck equation for the electron phase space density (PSD) $f_e$:
\begin{equation}
    \frac{\partial f_e}{\partial t} =
    \sum_{\mu'} \nabla_\mu \cdot
    (D_{\mu\mu'} \nabla_{\mu'} f_e) + S - L,
    \label{eqn:fokker_planck}
\end{equation}
where $\mu$ represents the adiabatic invariants $(M, J, \Phi)$, $D_{\mu\mu'}$ is the diffusion tensor encoding wave-particle interactions (chorus, EMIC, hiss, and ULF waves driving radial diffusion), $S$ represents injection from the plasma sheet, and $L$ represents losses via magnetopause shadowing and precipitation. The diffusion tensor $D_{\mu\mu'}$ depends on the wave power spectra and wave normal angle distributions, which are themselves functions of the magnetospheric state. This self-consistency makes the radiation belt problem one of the most computationally demanding in space physics.

Machine learning addresses this system along two operator paths. The first is \textit{flux propagation}: forecasting electron flux at specific operational locations from solar wind and geomagnetic index inputs, without solving the full PSD evolution. The second is \textit{wave-state inference}: recovering the wave populations that populate $D_{\mu\mu'}$ from spacecraft observations, which grounds the diffusion tensor in data rather than climatological averages. Physics-informed neural networks have shown that electron dynamics are better described by a drift-diffusion equation than by pure diffusion, and can recover effective diffusion coefficients directly from Van Allen Probes observations~\cite{2022JGRA..12730377C, camporeale2018machine}, suggesting that the functional form of $D_{\mu\mu'}$ itself can be learned from data.

Electron flux forecasting at GEO is one of the longest-standing ML applications in radiation belt physics, with feed-forward networks and LSTMs mapping $K_p$, solar wind speed, and flux history to MeV electron flux over lead times from hours to days. Recent work has extended this to pitch-angle-resolved differential flux at MEO altitudes, providing a richer characterization of the radiation environment relevant to GPS and Galileo orbits~\cite{2025SpWea..2304339W}. The connection between different energy populations has been exploited to improve operational forecasts: sub-relativistic electrons, which arrive at GEO earlier and are more continuously observed, serve as a precursor state from which relativistic flux can be predicted through a learned cross-energy propagation operator~\cite{2025PhFl...37b6618Y}. Transformer-based architectures that learn long-range temporal dependencies have improved upon LSTM baselines for GEO radiation forecasting, particularly during the growth and recovery phases of storms where the flux history over several days is informative~\cite{2024SpWea..2204119T}. Ensemble approaches combining multiple forecast members provide calibrated uncertainty bounds that are critical for satellite operations decisions~\cite{2026arXiv260515752K}. A more physically complete approach models the electron phase space density directly in adiabatic invariant coordinates: by first applying information theory to identify which solar wind and magnetospheric inputs carry the most information about PSD and at what lag times, neural networks trained on these physics-selected inputs produce a global $L^*$-resolved PSD model that can be driven purely by upstream solar wind observations~\cite{2022SpWea..2003090W}.

Populating $D_{\mu\mu'}$ with data-driven wave distributions is the complementary task. Identifying chorus, EMIC, hiss, and ULF wave intervals in spectrograms is a classification problem on high-dimensional, time-frequency data. Neural network models applied to Van Allen Probes wave data have generated global distributions of chorus wave occurrence and intensity spanning the full mission, supporting statistical studies of wave-driven diffusion that would be impractical with manual identification~\cite{2023SpWea..2103524C}. Prediction of plasmaspheric hiss spectral classes from upstream indices extends the wave-state estimate to times without direct wave measurements~\cite{2022FrASS...9.7801K}. Progress toward fully coupled radiation belt specification will require wave inference and flux propagation to share a common pipeline, with ML-predicted wave distributions feeding the diffusion tensor rather than being treated as independent outputs.

\subsection{Substorm dynamics and magnetotail reconnection}
\label{subsec:substorms}

A geomagnetic substorm is a transient episode in which magnetic energy accumulated in the magnetotail is rapidly released through reconnection, driving fast flows, magnetic field dipolarization, particle injection into the inner magnetosphere, and an intensification of the auroral oval. The magnetotail state during a substorm cycle may be written as
\begin{equation}
    \vec{S}_{\rm tail}(\vec{x},t) =
    \{\vec{B}_{\rm lobe},\, n_{\rm ps},\, T_{\rm ps},\,
      E_z^{\rm rec},\, \vec{v}_{\rm BBF},\, \Phi_{\rm rec}\},
    \label{eqn:tail_statevec}
\end{equation}
where $\vec{B}_{\rm lobe}$ is the lobe magnetic field (related to the stored energy via $B_{\rm lobe}^2/2\mu_0$), $n_{\rm ps}$ and $T_{\rm ps}$ are the plasma sheet density and temperature, $E_z^{\rm rec}$ is the reconnection electric field at the X-line, $\vec{v}_{\rm BBF}$ is the bursty bulk flow velocity, and $\Phi_{\rm rec}$ is the reconnection rate. This state passes through a characteristic loading-unloading cycle: during the growth phase, southward IMF drives flux transfer into the lobes and the plasma sheet thins; at onset, an instability rapidly releases the stored flux and returns it to the inner magnetosphere. The physical trigger of onset is a decades-long open problem, with multiple competing theories proposed for how and where in the magnetosphere-ionosphere system the instability is initiated. Because this transition is governed by a nonlinear current sheet instability regardless of which scenario applies, the timing of onset is not deterministically predictable from upstream conditions alone, which motivates probabilistic ML approaches across multiple observational channels.

The operator for substorm onset prediction maps the recent solar wind and index history to the probability of an onset within a future time window:
\begin{equation}
    \mathcal{P}_{\rm sub}:
    \{\vec{S}_{\rm sw}(t-\tau:t),\,
      \text{AL}(t-\tau:t),\,
      \vec{O}_{\rm aurora}(t)\}
    \rightarrow
    p(\text{onset} \mid t, t+\Delta t).
    \label{eqn:substorm_onset}
\end{equation}
Deep learning models applied to solar wind and AL index time series have demonstrated that the onset probability rises systematically during the growth phase and that the lead time over which the onset is predictable is intrinsically limited by the chaotic nature of current sheet thinning~\cite{2019SpWea..17.1534M}. Predicting the magnitude of the substorm response -- rather than just its timing -- requires addressing the severe imbalance between quiet-time and strong-event samples; imbalanced regression neural networks trained on the SuperMAG SML index have been shown to improve prediction of super-substorm-level electrojet intensifications that standard MSE-minimizing models systematically underestimate~\cite{2025SpWea..2304236C}. CNN-based classifiers applied to all-sky, DMSP, and ultraviolet auroral images exploit spatially resolved signatures -- equatorward oval expansion and onset arc brightening -- that in-situ indices cannot capture, and have been trained to recognize pre-onset morphology and identify onset time automatically~\cite{2023SpWea..2103300S, 2025JGRA..13033983H}. Pi2 magnetic pulsations, which are impulsively generated at substorm onset, provide a ground-based timing signature that has been exploited for automated onset detection using recurrent neural networks applied to magnetometer time series~\cite{2025IEEEA..13j9828A}.

Beyond onset timing, ML has begun to address the downstream consequences of reconnection in the tail. BBFs ($\vec{v}_{\rm BBF}$) are difficult to characterize statistically because only a handful of spacecraft sample the vast magnetotail at any time; ML models trained on THEMIS and MMS plasma sheet data have been developed to predict BBF characteristics from upstream and local plasma conditions~\cite{2025FrASS..1282607F}. At the reconnection site itself, determining where a spacecraft lies within the diffusion region geometry -- inflow, outflow, or separatrix -- is a classification problem that recurrent neural networks bridge between in-situ measurements and simulation geometry~\cite{2025JGRA..13034383W}. At the global scale, data mining of decades of magnetospheric magnetometer observations has revealed the X-line structure of magnetotail reconnection by assembling virtual spacecraft swarms from archived data~\cite{2023JGRA..12831066S}. Global substorm dynamics -- current sheet thinning, flux reconfiguration, and earthward transport -- have been reconstructed by combining low-altitude ELFIN precipitation measurements with data-mining-based magnetic field models, enabling a picture of the full tail evolution that no single in-situ mission can provide~\cite{2024SpWea..2204062S}. Together, these efforts extend ML coverage from the macroscopic onset trigger to the microscale dissipation region, connecting the full loading-unloading cycle of $\vec{S}_{\rm tail}$.


\section{Near-Earth space: ionosphere, thermosphere, and ground impacts}
\label{sec:nearearth}

Space weather exerts its most direct impact on technological systems through the near-Earth layers. Disturbances in the ionosphere degrade satellite navigation and communication, storm-time heating of the thermosphere accelerates orbital decay, and rapid geomagnetic variations at the ground drive currents through power grids. These layers are also the final link in the Sun-to-Earth chain, driven simultaneously from three directions. From above, the magnetosphere of Section~\ref{sec:magnetosphere} imposes field-aligned currents, particle precipitation, and convection electric fields; on the dayside, solar EUV and X-ray radiation ionizes and heats the system; and from below, upward-propagating tides and planetary waves perturb it. The result is a layered system of interacting state vectors, each acting as a boundary condition for the next and each only partially observable. Machine learning is both attractive and difficult here. Thousands of GNSS receivers, magnetometers, and radars operate continuously, but their coverage is spatially uneven and sparse, they measure the underlying state only indirectly, and the physics coupling the layers is strongly nonlinear. In the remainder of this section we move down through these layers, identifying for each its state vector, the operators that couple it to the regions above and below, the proxies on which operational models depend, and the machine learning methods used to infer, forecast, and emulate its state.

\subsection{The ionospheric state vector and the observational gap}
\label{subsec:iono_state}

The ionosphere is the partially ionized layer of the upper atmosphere spanning roughly 60 to 1000 km altitude. Its state vector can be written schematically as
\begin{equation}
    \vec{S}_{\rm ion}(\vec{x},t) =
    \{\, N_e,\, N_{i,s},\, T_e,\, T_i,\,
       \vec{v}_{\rm ion},\, \vec{E},\, \Sigma_P,\, \Sigma_H,\, \ldots \,\},
    \label{eqn:iono_statevec}
\end{equation}
where $N_e$ is the electron number density, $N_{i,s}$ the ion density for species $s$ (O$^+$, NO$^+$, O$_2^+$, H$^+$, He$^+$), $T_e$ and $T_i$ the electron and ion temperatures, $\vec{v}_{\rm ion}$ the ion drift velocity driven primarily by the electric field $\vec{E}$, and $\Sigma_P$ and $\Sigma_H$ the height-integrated Pedersen and Hall conductances that determine how the ionosphere closes magnetospheric currents. This state is governed by a set of coupled continuity, momentum, and energy equations for each species, with photoionization from solar EUV providing the primary source, recombination the primary loss, and transport redistributing plasma vertically and horizontally.

The ionospheric state is never directly observed in full. The primary observational access is through radio measurements. The total electron content (TEC) is the line-of-sight integral of $N_e$ along the ray path from a GNSS satellite to a ground or LEO receiver,
\begin{equation}
    \text{TEC}(t) = \int_{\rm path} N_e(\vec{x},t)\, ds,
    \label{eqn:tec_integral}
\end{equation}
and the vertical TEC (VTEC) is obtained by projecting the slant-path integral to the local vertical. The critical frequency $f_{\rm oF2}$ and the peak height $h_{\rm mF2}$ from ionosondes characterize the F2 layer maximum but provide no information about the electron density profile above the peak. Incoherent scatter radars recover full altitude profiles of $N_e$, $T_e$, $T_i$, and ion composition, but only at a sparse set of fixed locations. The observable vector is therefore
\begin{equation}
    \vec{O}[\vec{S}_{\rm ion}] =
    \{\text{TEC}_{\rm GNSS},\, f_{\rm oF2},\, h_{\rm mF2},\,
       N_e^{\rm ISR},\, S_4,\, \sigma_\phi,\, \ldots\},
    \label{eqn:iono_obs}
\end{equation}
where $S_4$ and $\sigma_\phi$ are the amplitude and phase scintillation indices that quantify the impact of small-scale electron density irregularities on trans-ionospheric radio signals.

The ionosphere can be viewed as the superposition of a background and a perturbation. Solar EUV and X-ray photoionization on the dayside establish the quiet-time background, influenced from below by upward-propagating tides and planetary waves. The storm-time perturbation is driven from above by the magnetosphere, described by the coupling operator
\begin{equation}
    \mathcal{P}_{\rm mag \to ion}: \vec{S}_{\rm mag}(t-\tau:t) \rightarrow
    \delta\vec{S}_{\rm ion}(\vec{x},t),
    \label{eqn:mag_to_ion}
\end{equation}
where field-aligned currents, particle precipitation, and convection electric fields modify the ionospheric electron density, temperature, and electrodynamics. Prediction is needed most during storms, when the perturbation becomes comparable to the background and the coupled system is at its most nonlinear.

\subsection{Global TEC forecasting and ionospheric specification}
\label{subsec:tec}

TEC is the most widely used summary of the ionospheric state because it is continuously observed by the global GNSS network and because it quantifies the ionospheric range delay that degrades positioning and navigation accuracy. The temporal propagation operator
\begin{equation}
    \mathcal{P}_{\rm TEC}: \{\text{TEC}(t-\tau:t),\,
    \vec{S}_{\rm proxy}(t)\}
    \rightarrow \text{TEC}(t+\Delta t)
    \label{eqn:tec_propagator}
\end{equation}
maps the recent history of TEC observations combined with solar and geomagnetic activity proxies to future TEC values. This is the dominant ML application in ionospheric research, with a broad diversity of architectures. LSTMs serve one-dimensional station-based forecasting, encoder-decoder ConvLSTMs predict global TEC maps, graph neural networks with transformer attention capture spatial dependencies across the irregular GNSS station network, and generative adversarial networks complete TEC maps in data-sparse regions~\cite{2020RemS...12.1354K, 2025IJSTA..18.1802L, 2022RemS...14.3547N}. A comprehensive review of deep learning architectures applied to this problem identifies ConvLSTM-based models as the current baseline for global map forecasting, while transformer architectures offer improved skill at longer lead times where temporal self-attention can integrate information across the previous 24 to 72 hours~\cite{2025RemS...17..124Z}.

Storm-time TEC prediction presents a harder version of this operator because the perturbation $\delta\text{TEC}$ driven by Eq.~\ref{eqn:mag_to_ion} can exceed the quiet-time background by factors of two to three, and the spatial pattern of the disturbance, which can be both positive (storm-enhanced density) and negative (main-phase trough), depends on the storm phase, local time, and season in ways that are not captured by global climatological models. CNN-BiLSTM architectures trained on decades of global ionospheric maps capture the onset and evolution of storm-time TEC disturbances, with geomagnetic indices as additional inputs measurably reducing the RMSE of long-lead forecasts~\cite{2024RemS...16.3160R, 2024SpWea..2203877R}. As throughout space weather forecasting, the underlying limitation is data imbalance. Extreme storms occur only a few times per solar cycle, so any training data set contains at most a handful of examples, and models trained on such imbalanced data are least reliable for the largest events, exactly those for which accurate TEC forecasts are most needed.

Beyond single-station or global map forecasting, the ionospheric specification problem is the inverse task of inferring the three-dimensional electron density distribution from TEC observations. Physics-based data assimilation systems such as IDA3D and GAIM have established the standard for three-dimensional $N_e(\vec{x},t)$ reconstruction, and their operational performance under storm conditions has been systematically documented~\cite{2005JGRA..11012306M}. Machine learning extends this assimilation framework in two directions. First, neural networks trained on large radio occultation datasets can produce three-dimensional electron density profiles with global coverage and at a time resolution impractical for iterative data assimilation algorithms~\cite{2021JASTP.22105702H, 2026AdSpR..78..344J}. Second, the IRI empirical model, which provides a climatological background for assimilation systems, can itself be replaced by a learned surrogate that corrects systematic biases identified from the long-term observation record~\cite{2020SpWea..1802411J, 2024JGeod..98...10G}.

\subsection{Plasma irregularities and ionospheric scintillation}
\label{subsec:scint}

Embedded within the large-scale ionospheric structure is a range of plasma irregularities with scale sizes from hundreds of kilometers to meters. These irregularities scatter radio waves, producing amplitude and phase fluctuations that degrade GNSS accuracy, disrupt satellite communications, and limit high-frequency radio propagation. The state vector of the irregular ionosphere extends $\vec{S}_{\rm ion}$ to include the electron density fluctuation spectrum:
\begin{equation}
    \vec{S}_{\rm irreg}(\vec{x},t) =
    \{\delta N_e(\vec{x},t),\,
       v_{\rm bubble},\, h_{\rm EPB},\, \ldots\},
    \label{eqn:irreg_statevec}
\end{equation}
where $\delta N_e$ is the electron density perturbation field, whose wavenumber spectrum spans the scales responsible for scintillation, $v_{\rm bubble}$ is the equatorial plasma bubble (EPB) drift velocity, and $h_{\rm EPB}$ is the bubble apex altitude. The EPB, the most consequential irregularity structure in the equatorial ionosphere, is a magnetic flux tube depleted of plasma that rises buoyantly through the background ionosphere during the post-sunset hours, persists for several hours, and generates irregularities across a wide range of scale sizes in its wake.

The scintillation prediction operator maps geophysical drivers and location to future scintillation indices:
\begin{equation}
    \mathcal{P}_{\rm scint}: \{\vec{S}_{\rm proxy}(t),\,
    \theta,\, \phi,\, t_{\rm LT},\, d\}
    \rightarrow
    \{S_4,\, \sigma_\phi\}(t+\Delta t),
    \label{eqn:scint_propagator}
\end{equation}
where $\theta$ and $\phi$ are the latitude and longitude, $t_{\rm LT}$ the local time, and $d$ the day of year. Random forest, XGBoost, and LSTM classifiers applied to ground-based GNSS scintillation records have demonstrated that the occurrence probability of strong scintillation is predictable from a combination of geomagnetic activity indices and local solar time, recovering the known physical drivers of the Rayleigh-Taylor instability that seeds EPBs without being explicitly trained on the instability criterion~\cite{2024Atmos..15.1213N}.  At high latitudes the relevant irregularities arise from different physics, with field-aligned structures seeded by auroral precipitation and polar-cap convection. There, stacked machine learning models that integrate multiple instrument types (GNSS, magnetometers, solar wind) improve on single-source scintillation predictors~\cite{2026JGRA..13134935Z}.

\subsection{Field-aligned currents, aurora, and high-latitude electrodynamics}
\label{subsec:fac_aurora}

Field-aligned currents (FACs) are the primary channel through which the magnetosphere deposits energy and momentum into the high-latitude ionosphere. The Region~1 and Region~2 FAC systems close through horizontal currents in the auroral ionosphere, driving the electrojet current system and producing the ground magnetic perturbations that underlie geomagnetic indices. The high-latitude ionospheric state vector is coupled to the magnetospheric state through
\begin{equation}
    \mathcal{P}_{\rm mag \to HL}:
    \{\vec{S}_{\rm mag},\, \vec{S}_{\rm sw}\}(t-\tau:t)
    \rightarrow
    \{J_{\rm FAC}(\theta,\phi,t),\,
       \Sigma_P(\theta,\phi,t),\,
       \Sigma_H(\theta,\phi,t),\,
       \Phi_{\rm pc}(t),\, \ldots\},
    \label{eqn:mag_to_hl}
\end{equation}
where $J_{\rm FAC}$ is the field-aligned current density as a function of magnetic latitude $\theta$ and longitude $\phi$, and $\Phi_{\rm pc}$ is the cross-polar-cap potential whose ML prediction was discussed in Section~\ref{subsec:sw_mag_coupling}. The physical complexity of this coupling is that the conductances $\Sigma_P$ and $\Sigma_H$ are set by the precipitation pattern, while the precipitation and the current closure in turn depend on those same conductances, a self-consistent nonlinear feedback that linearized empirical models handle poorly.

Aurora, the optical manifestation of particle precipitation, provides the spatially resolved observational access to this system that point measurements of FACs cannot supply. All-sky cameras capture the large-scale morphology of the auroral oval and the fine structure of auroral arcs, substorm bulges, and discrete emission features. Because expert labels exist for only a tiny fraction of the available imagery, self-supervised contrastive learning is particularly effective for this task. Adapting the SimCLR framework to learn representations of THEMIS all-sky images from unlabeled data, then fine-tuning a lightweight classifier on a small labeled subset, improves classification accuracy by almost ten percentage points over the previous benchmark and has enabled automatic annotation of the full THEMIS archive of roughly 700 million images into categories such as arc, diffuse, and discrete aurora~\cite{2021icml.conf..128J, 2024JGRH....1..292J}. This makes statistical characterization of auroral occurrence rates as a function of solar wind and geomagnetic conditions possible at a scale impractical with manual classification.

The particle precipitation that drives aurora also sets the ionospheric conductances $\Sigma_P$ and $\Sigma_H$ that appear in Eq.~\ref{eqn:mag_to_hl}. A particle precipitation model trained on decades of DMSP particle spectrometer data provides spatially resolved, near-real-time estimates of the electron energy flux and mean energy across the auroral zone, improving on climatological precipitation models that miss mesoscale variability~\cite{2021SpWea..1902684M}. An echo state network emulator of the REPPU global MHD simulation (SMRAI2) maps solar wind time series and dipole tilt angle to two-dimensional patterns of field-aligned current, ionospheric potential, and conductivity at five-minute temporal and approximately one-degree spatial resolution. It runs orders of magnitude faster than the parent simulation and enables real-time specification of the auroral current system~\cite{2024SpWea..2203720K}. A deep learning model trained on AMPERE observations predicts the global distribution of Birkeland (field-aligned) currents from a one-hour history of solar wind, IMF, and geomagnetic conditions, and captures the evolution of the current magnitude and latitudinal extent through storm onset, main phase, and recovery~\cite{2020JGRA..12527908K}.

These current specification products feed into high-latitude convection models that use ML to extend traditional empirical models (Weimer, Heppner-Maynard) to the regime of strong, variable driving where their statistical basis is sparse~\cite{2022SpWea..2002920B}. The quality of these convection maps depends on correctly classified radar returns. SuperDARN backscatter mixes ionospheric, ground, and sea scatter, and an ML classifier guided by ray-tracing physics reliably distinguishes these modes across the full network and removes a longstanding ambiguity in the derived convection patterns~\cite{2022SpWea..2003130K}. ML-AIM (Machine Learning based Auroral Ionospheric electrodynamics Model) brings these components together, chaining the AMPERE-trained FAC model with conductance models to solve the current continuity equation and recover the full ionospheric electric potential, electric fields, Pedersen and Hall currents, and Joule heating from a 60-minute history of solar wind, IMF, and geomagnetic indices. Validated against 3204 DMSP F17 polar crossings, ML-AIM outperforms the Weimer empirical model and SuperDARN-assimilated potentials for the cross-polar-cap potential, and correctly reproduces the two-cell convection pattern and its storm-time enhancement~\cite{2024SpWea..2203683G}.

A common caveat applies to the many models in this subsection that rely on DMSP data, whether for training or for validation. The DMSP spacecraft fly in sun-synchronous orbits that poorly sample the midnight sector, where substorm-driven precipitation is most dynamic, so models tied to these orbits inherit a systematic local-time gap. The SSJ detectors also degrade over each satellite's multi-year lifetime, and calibration differences across the fleet propagate directly into the learned models unless recalibrated databases with corrected fluxes and uncertainty estimates are used~\cite{2017JGRA..122.9056R}.

\subsection{The thermospheric state vector and solar EUV proxies}
\label{subsec:thermo_state}

The neutral thermosphere occupies altitudes of roughly 80 to 600 km and constitutes the ambient medium in which the ionosphere is embedded and through which low-Earth-orbit (LEO) satellites travel. Its fundamental state vector is
\begin{equation}
    \vec{S}_{\rm therm}(\vec{x},t) =
    \{\rho_n,\, T_n,\, \vec{v}_n,\,
      [{\rm O}],\, [{\rm N}_2],\, [{\rm O}_2],\,
      [{\rm He}],\, [{\rm H}],\, \ldots\},
    \label{eqn:therm_statevec}
\end{equation}
where $\rho_n$ is the total neutral mass density, $T_n$ the neutral temperature, $\vec{v}_n$ the neutral wind vector, and $[\text{O}]$, $[\text{N}_2]$, $[\text{O}_2]$, $[\text{He}]$, $[\text{H}]$ the number densities of the major species. The thermospheric state is set by a balance among solar EUV and soft X-ray heating, particle precipitation and Joule heating from the magnetosphere-ionosphere current system, advection by the neutral wind, molecular diffusion and thermal conduction, and upward-propagating energy from tides and planetary waves originating in the lower atmosphere.

The dominant global energy input to the thermosphere is the incident solar radiation spectrum, and its role is formalized by the operator
\begin{equation}
    \mathcal{P}_{\rm EUV \to therm}:
    \{\Phi_{\rm EUV}(\lambda,\, t-\tau:t),\, \ldots\}
    \rightarrow
    \vec{S}_{\rm therm}(\vec{x}, t),
    \label{eqn:euv_to_therm}
\end{equation}
where $\Phi_{\rm EUV}(\lambda, t)$ is the solar EUV and XUV irradiance spectrum as a function of wavelength $\lambda$ and time $t$, and the ellipsis denotes the non-radiative energy inputs, Joule heating, particle precipitation, and forcing from below, discussed above. The EUV flux above 10~nm determines the ionization and photodissociation rates that set the composition ratio $[\text{O}]/[\text{N}_2]$ and the total heating rate, while X-ray flux below 10~nm drives additional photoelectron heating during flares. This is the physically correct input to thermospheric models. However, continuous and spectrally resolved EUV measurements are not routinely available, because space-based EUV spectrometers have limited mission lifetimes, suffer calibration degradation, and are not always maintained operationally.

The gap between $\Phi_{\rm EUV}(\lambda,t)$ and operational forecasting is bridged by scalar solar activity proxies
\begin{equation}
    \vec{S}_{\rm proxy}(t) =
    \{F_{10.7},\, S_{10.7},\, E_{10.7},\, M_{10.7},\, Y_{10.7},\, A_p,\, K_p,\, \ldots\},
    \label{eqn:therm_proxies}
\end{equation}
where $F_{10.7}$ is the 10.7~cm microwave flux emitted by the solar corona, $S_{10.7}$ is an EUV index derived from SOHO/SEM 26--34~nm irradiance measurements, $E_{10.7}$, $M_{10.7}$, and $Y_{10.7}$ are additional EUV- and XUV-sensitive proxies, and $A_p$ and $K_p$ parameterize the geomagnetic component of thermospheric heating through Joule dissipation. The empirical density models NRLMSISE-00, JB2008, and DTM all operate on this proxy set rather than on $\Phi_{\rm EUV}(\lambda,t)$ directly. This substitution introduces a two-step removal from the fundamental physics. Proxies such as $F_{10.7}$ correlate with EUV irradiance statistically over solar rotations and cycles but do not track short-term EUV variability on timescales of hours, particularly during flares when soft X-rays can increase by orders of magnitude while $F_{10.7}$ changes little. The flare forecasting methods of Section~\ref{sec:flares} are therefore directly relevant here, since a predicted flare is in effect a forecast of the short-term EUV and X-ray input to Eq.~\ref{eqn:euv_to_therm} that the scalar proxies cannot represent. Machine learning inherits this proxy-level description and cannot transcend it unless EUV or X-ray observations are explicitly included as inputs. Machine learning can, however, extend the proxy record itself. LSTMs trained on sunspot number, sunspot area, and plage area have reconstructed the $F_{10.7}$, $S_{10.7}$, $M_{10.7}$, and $Y_{10.7}$ indices back to 1826, enabling model validation against extreme historical events that the modern index record does not capture~\cite{2026SpWea..2404757B}.

The coupling between the ionosphere and thermosphere adds a further layer. Because the neutral thermosphere and the ion-electron plasma are co-located and exchange momentum through ion-neutral collisions, the thermospheric wind drives ion drift at high latitudes and the ionospheric electric field drives neutral wind through the same collision mechanism. This two-way coupling is expressed as
\begin{equation}
    \mathcal{P}_{\rm IT}:
    \{\vec{S}_{\rm ion}(t),\, \vec{S}_{\rm therm}(t)\}
    \rightarrow
    \{\vec{S}_{\rm ion}(t+\delta t),\, \vec{S}_{\rm therm}(t+\delta t)\},
    \label{eqn:it_coupling}
\end{equation}
and is what full physics-based ionosphere-thermosphere (IT) models such as TIEGCM and GITM solve. Machine learning applied to TIEGCM simulation output has demonstrated that the neutral wind and electric field profiles hidden within the model can be inferred from the electron density distribution alone, providing a neural inverse model that recovers thermospheric state variables from the more routinely observed ionospheric quantities~\cite{2026SciA...12a2406C}.

\subsection{Neutral density prediction and satellite drag}
\label{subsec:neutral_density}

The operationally critical projection of $\vec{S}_{\rm therm}$ is $\rho_n$ at satellite altitude, because it determines the atmospheric drag force that drives orbit decay and introduces uncertainty into conjunction assessment for space debris avoidance. The problem grows more pressing every year as commercial mega-constellations such as Starlink add thousands of satellites to low Earth orbit. The stakes were made clear in February 2022, when a minor geomagnetic storm enhanced thermospheric density enough to cause the loss of dozens of newly launched Starlink satellites to atmospheric drag~\cite{2022SpWea..2003193F, 2023SpWea..2103330B}. The neutral density varies by factors of two to ten between solar minimum and maximum, and by similar factors within hours during geomagnetic storms. The propagation operator for satellite drag forecasting can be written as
\begin{equation}
    \mathcal{P}_{\rm drag}:
    \{\vec{S}_{\rm proxy}(t-\tau:t+\Delta t_{\rm fcast})\}
    \rightarrow
    \rho_n(h_{\rm sat}, \theta, \phi, t+\Delta t),
    \label{eqn:drag_propagator}
\end{equation}
where the input reaches into the forecast window because the future proxies must themselves be predicted. This double uncertainty, in the proxies and in the density response to them, sets the practical limit on drag prediction skill.

The empirical models in current operation (NRLMSISE-00, JB2008, and the DTM series) map $\vec{S}_{\rm proxy}$ to $\rho_n$ through fitted analytic functions~\cite{2026JAnSc..73...32M}. Machine learning contributes to this pipeline in four ways, replacing the empirical models with fast surrogates, improving the density estimate itself, forecasting the proxy inputs, and quantifying the drag-coefficient uncertainty that density alone does not capture.

First, ML surrogates of operational density models offer faster inference and can be embedded in orbit propagation codes running at the speed required for operational conjunction screening~\cite{2022SpWea..2002915L}. MSIS-UQ replaces the exospheric temperature parameterization in NRLMSIS~2.0 with an ML model that outputs a calibrated probability distribution rather than a point estimate. It reduces the difference between model and satellite-derived densities by 25\%, is statistically 11\% closer to satellite densities than the operational High Accuracy Satellite Drag Model (HASDM), and captures post-storm thermospheric overcooling that NRLMSIS~2.0 alone misses~\cite{2022SpWea..2003267L}.

Second, machine learning improves the density estimate itself, moving beyond the analytic form of the empirical models. One route emulates physics-based IT models. Building on reduced-order frameworks for calibrating physics-based IT models~\cite{2018SpWea..16.1086M}, the ROPE (Reduced Order Probabilistic Emulator) framework applies PCA to compress the TIEGCM output space and LSTM networks to propagate the reduced state forward in time, reproducing storm-time density with orbital position errors below 5~km for the November 2003 extreme storm at a fraction of the parent model's cost~\cite{2023SpWea..2103345L}. A second route trains directly on satellite-derived densities from the CHAMP, GOCE, GRACE, and Swarm missions and corrects the storm-time enhancements that NRLMSISE-00 and JB2008 underestimate by factors of two or more. KARMAN is an open-source framework that ingests precise orbit determination-derived densities together with solar irradiance and geomagnetic inputs and reduces the mean absolute percentage error from the 40--60\% range of empirical models down to approximately 20\%~\cite{2024SpWea..2203652A}. Hybrid frameworks that embed the empirical NRLMSIS model as prior knowledge within a deep network converge faster in training and generalize more reliably to satellites outside the training data than purely data-driven networks~\cite{2025Atmos..16..539L}.

Third, the full drag forecast requires predicting $\vec{S}_{\rm proxy}$ itself, since the density response depends on solar and geomagnetic inputs not yet known at issue time. Neural network ensembles applied jointly to $\{F_{10.7},\, S_{10.7},\, M_{10.7},\, Y_{10.7}\}$ improve over the operational USAF linear forecast by 12--18\% in RMSE and provide the first calibrated probabilistic forecasts for the full JB2008 driver set~\cite{2023SpWea..2103675D, 2024SpWea..2203785D}.

Fourth, the drag coefficient $C_D$, which depends on satellite shape, attitude, and the gas-surface interaction, is an independent source of uncertainty that density models do not address; stochastic neural network models for $C_D$ trained on numerical gas-surface interaction simulations predict both its mean and its uncertainty across the full range of satellite attitudes, and have been used to quantify the impact of drag-coefficient uncertainty on orbit prediction and space traffic management~\cite{2023AdSpR..72..922P}. Tying all four together, fully data-assimilative, physics-based density specification that maps real-time solar and geomagnetic observations into the forcing terms of IT models offers a promising path toward storm-time accuracy at substantial computational cost~\cite{2018SpWea..16..736S, 2018SpWea..16.1086M}, and provides a reference point for faster ML approaches. In the same direction, the SWAMI (Space Weather Atmosphere Models and Indices) project blended the DTM empirical model with the Met Office whole-atmosphere model and introduced the higher-cadence Hp30 and Hp60 geomagnetic indices, with neural networks applied to forecasting these indices for improved storm-time density modeling~\cite{2020JSWSC..10...18J}.

\subsection{Geoelectric field and geomagnetically induced currents}
\label{subsec:gic}

The ground-level consequence of the magnetospheric storm is a time-varying magnetic field perturbation $\delta\vec{B}(\vec{r},t)$ that by Faraday's law induces a geoelectric field at the conducting surface of the Earth. The induced field is related to the surface magnetic perturbation through a convolution with the Earth's impedance kernel,
\begin{equation}
    \vec{E}_{\rm geo}(\vec{r},t) =
    \frac{1}{\mu_0} \int_{0}^{\infty}
    \mathcal{Z}_\sigma(\vec{r},\tau)\,
    \delta\vec{B}(\vec{r},\, t-\tau)\, d\tau,
    \label{eqn:geoelectric}
\end{equation}
where $\mathcal{Z}_\sigma$ is the Earth's transfer function (impedance tensor) determined by the subsurface electrical conductivity structure $\sigma(\vec{r})$, and the relation is commonly evaluated in the frequency domain as $\vec{E}_{\rm geo}(\omega) = \mathcal{Z}_\sigma(\omega)\, \delta\vec{B}(\omega)/\mu_0$. This geoelectric field drives geomagnetically induced currents (GICs) in long conducting infrastructure (power transmission lines, pipelines, railway systems, and submarine cables) through
\begin{equation}
    \mathcal{P}_{\rm GIC}:
    \vec{E}_{\rm geo}(\vec{r},t)
    \rightarrow
    I_{\rm GIC}(\vec{r}_k,\, t),
    \label{eqn:gic_propagator}
\end{equation}
where $\vec{r}_k$ are the network nodes, and the mapping from the geoelectric field to the actual current in each network element depends on the grid topology, line impedances, and transformer characteristics. GICs can saturate power transformers, increase reactive power demand, cause voltage instability, and in extreme cases contribute to transformer damage. The 1989 Quebec blackout is the canonical auroral-zone example~\cite{2019SpWea..17.1427B}, while the 2003 Halloween storms~\cite{2005SpWea...3.8C03P} and the May 2024 Gannon storm~\cite{2025SpWea..2304235C} demonstrated that moderate to severe geomagnetic storms produce operationally significant GICs at mid-latitudes as well, not only at the auroral-zone locations where $\delta\vec{B}$ is largest.

Machine learning addresses the GIC prediction chain at two points. The first is forecasting $\delta\vec{B}$ or $d\vec{B}/dt$ at ground level from upstream solar wind observations, which replaces the full physical chain from solar wind to magnetosphere to ionosphere to ground induction with a learned end-to-end mapping:
\begin{equation}
    \mathcal{P}_{\rm \delta B}:
    \vec{S}_{\rm sw}(t-\tau:t)
    \rightarrow
    \delta\vec{B}(\vec{r},\, t+\Delta t).
    \label{eqn:deltab_propagator}
\end{equation}
Deep learning models applied to the ground magnetic perturbation forecasting problem have demonstrated that convolutional-recurrent architectures outperform persistence and linear baselines for the largest perturbation events, which are the most operationally critical~\cite{2022SpWea..2003045U, 2022FrASS...931407S, 2020FrASS...7...72K}. A deep Gaussian process model, GeoDGP, provides global probabilistic forecasts of the horizontal perturbation at arbitrary locations with one-hour lead time, and outperforms both the operational physics-based Michigan Geospace model and earlier data-driven models across 24 storms, including the May 2024 Gannon storm~\cite{2025SpWea..2304301C}. Probabilistic forecasting of $d\vec{B}/dt$ spikes at mid-latitude stations addresses the direct driver of GIC transients, and ensemble quantile regression models have been shown to provide calibrated estimates of the probability that operational thresholds will be crossed~\cite{2023SpWea..2103446C}. A gray-box approach that combines a physically motivated functional form for the ground response with ML-fitted coefficients provides better interpretability than a fully black-box model while retaining the nonlinear expressiveness needed to capture the storm-time amplification~\cite{2020JGRA..12527684C}. Related work revisits which benchmark metrics best evaluate model skill for this rare-event forecasting problem~\cite{2022FrASS...9.9740P}.

The second point is the geoelectric field estimation step in Eq.~\ref{eqn:geoelectric}. Given $\delta\vec{B}$, computing $\vec{E}_{\rm geo}$ requires knowledge of $\sigma(\vec{r})$, which is heterogeneous and not uniformly well-characterized. Data-driven transfer function models trained on simultaneous magnetometer and geoelectric field measurements avoid the conductivity inversion problem entirely, learning the local $\vec{B} \to \vec{E}_{\rm geo}$ mapping from instrument pairs~\cite{2025SpWea..2304180M}. LiveWire, a multi-fidelity ML system for horizontal geoelectric field prediction at one-hour lead time, incorporates both physics-based induction computations and data-driven corrections to handle the spatial variability of the subsurface conductivity response~\cite{2024JGRH....1..151H}. LSTM models trained on historical GIC records from power grid monitoring have shown that direct $\vec{S}_{\rm sw} \to I_{\rm GIC}$ mappings are feasible at specific well-instrumented sites. The entire physical chain is compressed into a learned end-to-end operator, at the cost of portability to unmonitored sites~\cite{2022SpWea..2002907B}. The physical limit on GIC predictability is set by the chaotic, fine-scale structure of $\delta\vec{B}$ during substorm expansions (Section~\ref{subsec:substorms}), which are the intervals of highest $dB/dt$ and therefore highest GIC risk. Establishing the predictable horizon for $I_{\rm GIC}$ under different driving conditions remains an open question with direct operational relevance to grid resilience planning.

\section{Conclusion and future directions}
\label{sec:conclusion}

In this review, we have treated space weather as a coupled Sun-to-Earth state-estimation and forecasting problem. The state vector formalism defines the physical object that {\ml} is attempting to infer, propagate, emulate, or classify. In this language, {\infer} maps partial observations to an estimate of the hidden state, $\mathcal{P}_{a\to b}$ transfers information between coupled subsystems or advances a state in time, and {\emulate} replaces expensive numerical models with fast surrogates. Such a formalism grounds the {\ml} application on the physics rather than on the architecture of the {\ml} solution. The same transformer, LSTM, random forest, graph network, or Gaussian process has different scientific meaning depending on whether it is reconstructing $\vec{S}_{\rm sw}$, forecasting a flare probability, compressing $\vec{S}_{\rm mag}$ into an index, or estimating a geoelectric-field impact.

The salient conclusions that emerge from such a study are:
\begin{itemize}
    \item {\bf The solar wind and heliosphere}: {\ml} has entered nearly every stage of solar-wind modeling, from optimization of empirical coronal boundary prescriptions, to direct coronal-image forecasting, to in-situ autoregression, to fast surrogates of global MHD and kinetic simulations. These methods are valuable because they reduce the computational cost of heliospheric context and make ensemble forecasting more feasible. However, the dominant limitation is not only point-wise error in speed or density. A single spacecraft observes a one-dimensional trajectory through a three-dimensional, evolving plasma, while global models depend on stale magnetograms, missing far-side information, simplified heating prescriptions, and uncertain source-surface mappings. Conscious strategies need to be adopted to bridge this observation gap, by potentially including known physics.
    \item {\bf Flares, CMEs, and SEPs}: Eruptive-event forecasting is the clearest example where the operational target is a coupled probability distribution, not a single label. Solar flares, CMEs, and SEP events share source-region information but differ in onset, association rules, transport physics, and impact variable. Across this literature, the practical challenge is not only prediction accuracy but coherent event definition. Different catalogs use different thresholds, association rules, start and end times, and treatment of overlapping events. These choices change the labels that the model sees during training. A network trained on one catalog may therefore learn a mixture of solar physics, instrument sensitivity, and human annotation rules. This is especially consequential for rare high-impact events, where a small number of examples can dominate the learned decision boundary. The most useful eruptive-event models should therefore report calibrated probabilities, uncertainty, event timing, and physically interpretable drivers. In the language of Eq.~(\ref{eqn:eruptive_operator}), the goal is not merely to classify a historical archive, but to estimate the evolving eruptive state well enough that downstream Sun-to-Earth predictions become more reliable.
    \item {\bf The magnetosphere}: The magnetosphere demonstrates both the power and the danger of compressed observables. Forecasting $K_p$, Dst, SYM-H, AE, or AL is operationally important, but these indices are projections of the magnetospheric state rather than the state itself. A model that predicts an index may learn the composition of solar-wind driving, magnetospheric dynamics, ground-station coverage, and index construction. At the same time, {\ml} has made important progress in learning solar wind--magnetosphere coupling, identifying plasma regions and boundaries, emulating ring-current and global simulations, forecasting radiation belt fluxes, and detecting substorm onset. The next step is to connect these outputs more explicitly to physical state variables such as particle phase space density, current systems, wave power, reconnection geometry, and field-aligned current closure. This is especially important during strong driving, where proxy measurements and L1 propagation errors can be mistaken for physical nonlinearities in the learned operator.
    \item {\bf Near-Earth impacts}: In the ionosphere, thermosphere, and ground environment, {\ml} is closest to the technological consequences of space weather. Global TEC forecasting, scintillation prediction, auroral image classification, field-aligned current specification, neutral-density and satellite-drag forecasting, and geoelectric-field or GIC prediction all operate on dense but uneven observational networks. The central issue is that the variables most relevant to users are often not the directly observed quantities. TEC is an integral of electron density, drag depends on neutral density and satellite properties, and GIC depends on magnetospheric currents, ground conductivity, and infrastructure topology. Hence, a visually plausible TEC map, a low-RMSE density model, or an accurate local magnetic perturbation forecast may still fail in the final impact variable. Validation should therefore be performed in physics and impact space, with uncertainty carried through each step of the chain.
    \item {\bf Operational forecasting}: Across all subsystems, the dominant obstacles are partial observability, non-stationarity, rare extremes, latency, and heterogeneous labels. The system changes across the solar cycle, instruments degrade or are replaced, spacecraft sample different regions of state space, and operational pipelines change the available data stream. Random train-test splits are therefore insufficient, because adjacent samples from the same active region, solar rotation, storm, or instrument era are not independent. Forecast systems should be validated across time, activity level, mission epoch, and event severity, with all-clear performance and the cost of false alarms v/s missed events reported explicitly.
\end{itemize}

These points lead to a practical agenda for the next phase of {\ml} in space weather:
\begin{itemize}
    \item Move from point estimates to probabilistic operators with calibrated posteriors, explicit failure modes, and event-dependent uncertainty. In a partially observed Sun-to-Earth system, a deterministic forecast can hide the cases where the model is extrapolating beyond the training distribution.
    \item Prioritize hybrid physics-{\ml} formulations that encode conservation laws, symmetries, geometry, and known causal pathways. Purely data-driven models are useful, but the most credible operational systems will combine physical models, data assimilation, learned surrogates, and uncertainty propagation. This is already incorporated to a significant extent in {\ml} for  magnetospheric / ionospheric physics.
    \item Design models and datasets for non-stationarity. Training and evaluation should account for changing solar-cycle phase, instrument calibration, spacecraft sampling, operational latency, and evolving event definitions. A model that performs well only within one catalog, mission, or activity regime has limited physical value. 
    \item Treat interpretability as a physical requirement. Learned features should be testable against known mechanisms such as open-flux expansion, active-region non-potentiality, magnetic reconnection, ring-current injection, wave-particle diffusion, Joule heating, and ground induction. 
    \item Build common benchmark datasets and reporting standards that preserve event metadata, issue time, data availability, label provenance, uncertainty, and operational cost. This is especially important for rare extremes, where small differences in catalog construction can change the apparent skill of a model.
\end{itemize}

The goal is no longer to show that {\ml} can emulate a particular task on a curated archive. The goal is to establish which learned operators remain physically valid under distribution shift, sparse supervision, incomplete observability, and operational latency. A mature Sun-to-Earth forecasting system would allow different models to exchange uncertainty-aware representations of the state vector, so that the solar atmosphere, heliosphere, magnetosphere, ionosphere, thermosphere, and ground-impact layers are not treated as disconnected prediction problems. This transition toward coupled, probabilistic, physics-aware {\ml} is essential if data-driven space-weather modeling is to move from retrospective skill to reliable operational decision support.


\begin{acknowledgement}
VU would like to acknowledge NASA for support under award number 80NSSC25K7956 in support for writing this chapter. The authors acknowledge the use of AI tools for data collection and compilation. The agentic AI pipeline is made available on Github.  The views expressed are those of the author and do not necessarily reflect the official policy or position of the Department of the Air Force, the Department of Defense, or the U.S. government. Approved for public release AFRL NUMBER TBD
\end{acknowledgement}



\bibliographystyle{plain}
\bibliography{sun_to_earth,bashi_refs_export}


\end{document}